\documentclass[aps,prb,twocolumn,superscriptaddress,floatfix,showpacs,10pt,letterpaper]{revtex4}

\pdfoutput=1

\usepackage{graphicx}
\usepackage{dcolumn}
\usepackage{bm}
\usepackage{amsfonts}
\usepackage{amsmath}
\usepackage{multirow}

\begin{document}

\begin{titlepage}

\title{Classification of Gapped Symmetric Phases
in 1D Spin Systems}

\author{Xie Chen}
\affiliation{Department of Physics, Massachusetts Institute
of Technology, Cambridge, Massachusetts 02139, USA}

\author{Zheng-Cheng Gu}
\affiliation{Kavli Institute for Theoretical Physics,
University of California, Santa Barbara, CA 93106, USA}

\author{Xiao-Gang Wen}

\affiliation{Department of Physics, Massachusetts Institute
of Technology, Cambridge, Massachusetts 02139, USA}
\affiliation{Institute for Advanced Study, Tsinghua
University, Beijing, 100084, P. R. China}

\begin{abstract}
Quantum many-body systems divide into a variety of
phases with very different physical properties. The question
of what kind of phases exist and how to identify them seems
hard especially for strongly interacting systems. Here we
make an attempt to answer this question for gapped
interacting quantum spin systems whose ground states are
short-range correlated. Based on the local unitary
equivalence relation between short-range correlated states
in the same phase, we classify possible quantum phases for
1D matrix product states, which represent well the class of
1D gapped ground states. We find that in the absence of any
symmetry all states are equivalent to trivial product
states, which means that there is no topological order in
1D. However, if certain symmetry is required, many phases
exist with different symmetry protected topological orders.
The symmetric local unitary
equivalence relation also allows us to obtain some simple
results for quantum phases in higher dimensions when some symmetries are present.
\end{abstract}


\maketitle

\vspace{2mm}

\end{titlepage}

\bibliographystyle{apsrev}

\section{Introduction}

For a long time, we believed that Landau symmetry-breaking
theory\cite{L3726,LanL58} describes all possible orders in
materials, and all possible (continuous) phase transitions.
However, in last twenty years, it has become more and more
clear that Landau symmetry-breaking theory does not describe
all possible orders. For example, different fractional
quantum Hall (FQH) states\cite{TSG8259,L8395} all have the
same symmetry. Thus it is impossible to use symmetry
breaking to characterize different FQH states.

If Landau symmetry breaking theory is not enough, then what
should we use to describe those new states of matter? It
turns out that we need to develop a totally new theory, to
described the new types of orders -- topological/quantum
order\cite{Wtoprev,Wqoslpub} -- that appear in the FQH
states and the spin liquid states.

In \Ref{LWstrnet}, a systematic understanding of a large
class of topological orders in strongly correlated bosonic
systems without symmetry has been developed based on
string-net condensations.  In \Ref{CGW1035}, the string-net
classification of topological orders was generalized, based
on local unitary (LU) transformations.  In
\Ref{Wqoslpub,KLW0800,GW0969}, topological orders with
symmetry are studied using projective symmetry group and
tensor network renormalization.  But so far, we still do not
have a complete classification of topological orders for
interacting systems.

Recently, for non-interacting gapped fermion systems with
certain symmetries, a complete classification of topological
phases has been developed based on the
K-theory.\cite{K0986,RSF0957} A generalization of the free
fermion result to interacting cases has been obtained for 1D
systems.\cite{FK1009}

For 1D bosonic systems, \Ref{HKH0831a} studied quantized
Berry phases for spin rotation symmetric systems.
\Ref{PBT1039} studied the entanglement spectrum and the
symmetry properties of matrix product states.  Using those
tools, they obtained some interesting results which are
special cases of the situation considered here.

In this paper, we will apply the approach used in
\Ref{VCL0501,CGW1035} to 1D strongly correlated systems.  We
will combine the local unitary transformation with the
symmetry properties of matrix product states, and try to
obtain a complete classification of all gapped phases
in 1D quantum spin systems with certain simple
symmetries.  We find that
\begin{enumerate}
\item[(a)]
\emph{if there is no symmetry (including translation
symmetry), all gapped 1D spin systems belong to the same
phase.}
\item[(b)]
\emph{for 1D non-translation invariant (NTI) spin systems with ONLY
an on-site symmetry described by a group $G$, all
the phases of gapped systems that do not break the symmetry
are classified by the equivalence classes in the second cohomology group $H^2(G,
\C)$ of the group $G$, provided that the physical states on
each site form a linear representation of the group
$G$.}
\end{enumerate}
(Note that the equivalence classes in $H^2(G,
\C)$ classify the types of projective representations of
$G$ over the field $\C$ of complex numbers. Appendix \ref{prorep} gives a brief introduction on projective representation and the second cohomology group.) In certain cases where $G$ has infinitely many 1D representations, for example when $G=U(1)$, further classification according to different 1D representations exist.

But quantum states are defined only up to global change of phases, therefore the symmetry operations of group $G$ only needs to be represented by operators $u(g)$ which satisfy
\begin{equation}
u(g_1)u(g_2)=e^{i\theta(g_1,g_2)}u(g_1g_2)
\end{equation}
for any group element $g_1$ and $g_2$. Such operators form a
projective representation of group $G$. When we consider
this general case, we find that the classification result
remains the same as with linear representations, that is,
\begin{enumerate}
\item[(c)] \emph{for 1D non-translation invariant (NTI) spin systems with
ONLY an on-site symmetry described by a group $G$,
all the phases of gapped systems that do not break the symmetry
are classified by the equivalence classes in the second cohomology group $H^2(G,
\C)$ of the group $G$, provided that the
physical states on each site form a projective
representation of the group $G$.}
\end{enumerate}
In certain cases where $G$ has infinitely many 1D representations, further classification according to different 1D representations exist.

Applying these general results to specific cases allows us to reach the following conclusions: First, result (a) means that there is no non-trivial
topological order in 1D systems without any symmetry; Using
result (b), we find that NTI integer-spin chain with
only on-site $SO(3)$ spin rotation symmetry can have two and
only two different phases that do not break the $SO(3)$
symmetry.  Result (c) implies that NTI half-integer-spin
chain with only on-site $SO(3)$ spin rotation symmetry (which
is represented projectively) also have two and only two
gapped phase that does not break the $SO(3)$ symmetry; We
note the cyclic $\Z_n$ group has no non-trivial projective
representations, thus NTI spin chain with only on-site $\Z_n$
symmetry can have one and only one gapped phase that does
not break the $\Z_n$ symmetry; The $U(1)$ symmetry group has no non-trivial projective representation either, however, due to the special structure of the group of 1D representations of $U(1)$, NTI spin chain with only on-site $U(1)$
symmetry can have three and only three gapped phase that does
not break the $U(1)$ symmetry.

We also considered systems with translation invariance (TI) and correspondingly many results have been obtained.
\begin{enumerate}
\item[(a)] \emph{if there is no other symmetry, all gapped TI systems
belong to the same phase.} (This has been discussed as the generic case in \Ref{VCL0501}).
\item[(b)] \emph{for 1D spin systems with
ONLY translation symmetry and an on-site symmetry described by a group $G$,
all the phases of gapped systems that do not break the two
symmetries
are labeled by the equivalence classes in the second cohomology group $H^2(G,
\C)$ of the group $G$ and different 1D representations $\alpha(G)$ of $G$, provided that the
physical states on each site form a linear
representation of the group $G$.}
\end{enumerate}

Similar to the NTI case, we should consider projective representations of $G$ at each site. However, we find that,
\begin{enumerate}
\item[(c)] \emph{there is no translation invariant gapped ground
state symmetric under on-site symmetry of group $G$ which is
represented projectively on the state space at each site.}
\end{enumerate}

In particular, we can show that the $SO(3)$ spin rotation
symmetric integer spin chain has two different gapped TI
phases:\cite{BTG0819,GW0931,PBT0959} the spin-0 trivial
phase and the Haldane phase\cite{H8364}, while translation
invariant $SO(3)$ symmetric half-integer spin chain must
either be gapless or have degeneracy in ground space due to
broken discrete symmetries\cite{LSM6107,HKH0831}. On the
other hand, the $SU(2)$ symmetric spin chains where on-site
degrees of freedom contain both integer-spin and
half-integer-spin representations have only one gapped TI
phase. We also show that the spin chain with only
translation and parity symmetry (defined as exchange of sites together with an on-site $Z_2$ operation) has four different gapped TI phases.\cite{BTG0819,GW0931,PBT0959}

For systems with time reversal symmetry, we find that NTI time reversal symmetric systems belong to two phases while TI time reversal symmetric phases on integer spin systems have two phases and those on half-integer spin systems are either gapless or have degeneracy in ground space.

The paper is organized as follows: section \ref{qphase}
gives a detailed definition of gapped quantum phases and
explains how that gives rise to an equivalence relation
between gapped ground states within the same phase; section
\ref{1DGapMPS} shows that short-range correlated matrix product states represents faithfully 1D gapped ground states and hence will be our object of study; section \ref{notop} discusses the situation where no symmetry is required and found no topological order in 1D; section \ref{symtop} gives the classification of phases for 1D systems with certain symmetries, for example on-site symmetry and time reversal symmetry. It classifies phases in translational invariant systems, and furthermore in systems where translational invariance is present together with other symmetries such as on-site symmetry, parity symmetry and time reversal symmetry;
section \ref{highD} generalizes some simple 1D results to
higher dimensions;
finally in section \ref{conc} we summarize our results and conclude this paper.

\section{Definition of Quantum Phases}
\label{qphase}
\begin{figure}
\begin{center}
\includegraphics[scale=0.5]{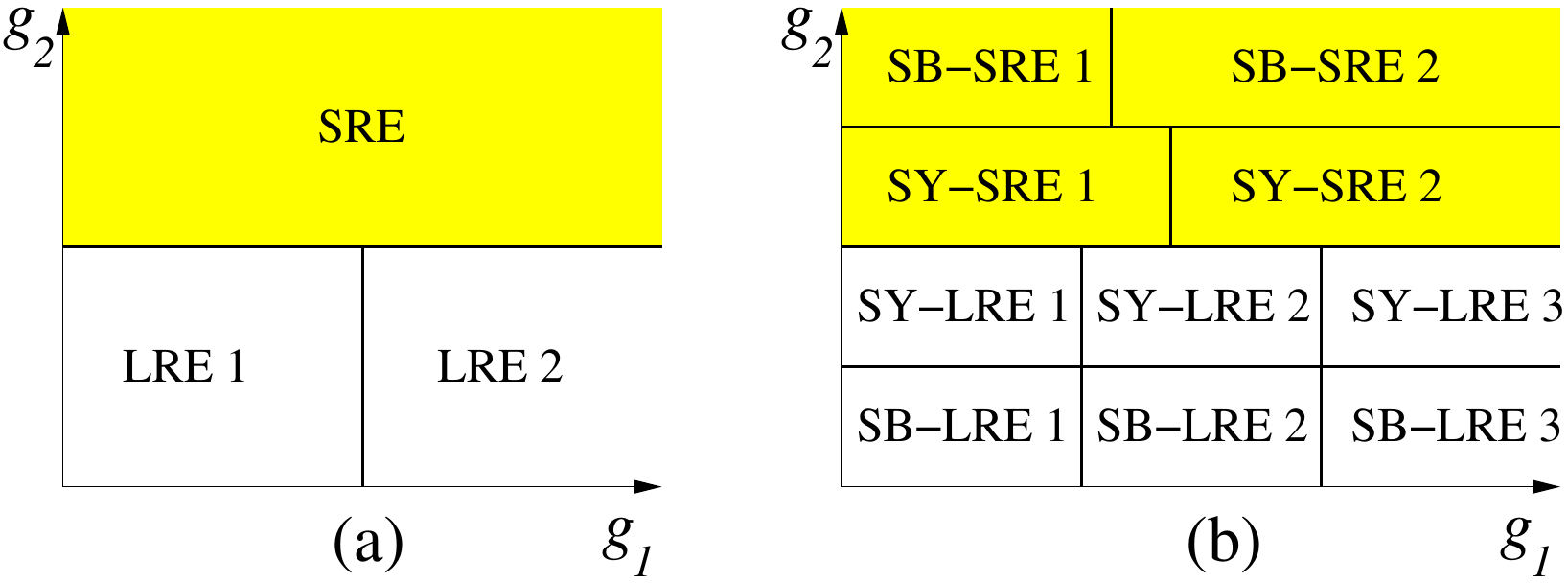}
\end{center}
\caption{
(Color online)
(a) The possible phases for class of Hamiltonians $H(g_1,g_2)$ without
any symmetry restriction.
(b) The possible phases for class of Hamiltonians $H_\text{symm}(g_1,g_2)$ with
some symmetries.
The shaded regions in (a) and (b) represent the phases with
short range entanglement.}
\label{fig:phase}
\end{figure}

To obtain the above stated results, we need to first briefly
discuss the definition of quantum phases. A more detailed
discussion can be found in \Ref{CGW1035}. Quantum phase describes
an equivalence relation between quantum systems. Systems we consider live on a $n$-dimensional lattice and interactions are local (with finite range). A gapped quantum
phase is usually defined as a class of gapped Hamiltonians
which can smoothly deform into each other without closing
gap and hence without any singularity in the local properties of
ground state. Such an equivalence relation between
Hamiltonians can be reinterpreted as an equivalence relation
between ground states, as discussed in \Ref{VCL0501,CGW1035}:
\textit{Two gapped ground states belong to the same phase if
and only if they are related by a local unitary (LU)
transformation}. \footnote{While it might seems insufficient to discuss equivalence between Hamiltonians completely in terms of equivalence between ground states, as a local unitary transformation mapping $|\phi_1\rangle$ to $|\phi_2\rangle$ might not map the corresponding Hamiltonian $H_1$ to $H_2$, but instead to some $H_2'$. However, as $H_2$ and $H_2'$ are both gapped and have the same ground states, their equivalence is obvious.}
As LU transformations can change
local entanglement structure but not the global one, states
in the same phase have the same long range entanglement (LRE)
and hence the same topological order.\cite{Wtoprev,Wen04}
States equivalent to product states have only short range
entanglement (SRE) and hence trivial topological order. All
the states with short range entanglement belong to the same
phase while states with long range entanglement can belong
to the different phases. These considerations lead to the
phase diagram as illustrated in Fig. \ref{fig:phase}(a), for the class of
systems without any symmetry requirement.

A LU transformation $U$ can either take the form
of finite time evolution with a local Hamiltonian
\begin{align}
U = \mathcal{T}[e^{-i\int_0^1 dg\, \tilde{H}(g)}]
\label{LU_H}
\end{align}
 where
$\mathcal{T}$ denotes time ordered integral and
$\tilde{H}(g)$ is a sum of local Hermitian terms. Or $U$ can take the
form of a constant depth quantum circuit
\begin{align}
U=\prod_{i_1} U^{(1)}_{i_1}...\prod_{i_R} U^{(R)}_{i_R}
\label{LU_U}
\end{align}
which is composed of $R$ layers of unitaries and the
$U^{(k)}_{i_k}$'s within each layer $k$ are local and
commute with each other.\cite{CGW1035} These two forms of
LU transformations are equivalent to each other
and we will mainly take the quantum circuit form for the discussion
in this paper (the time evolution form is used for discussion of translation invariant systems). More generally, we will consider equivalence
relation between states defined on different Hilbert spaces
and hence we allow a broader notion of unitarity (which is
called generalized LU transformation in
\Ref{CGW1035} and corresponds to the disentanglers and isometries in MERA\cite{Vidal0709}). Each LU operation $U^{(k)}_{i_k}$ we consider
in the quantum circuit will act unitarily on the support
space of the reduced density matrix of the region with
$(U^{(k)}_{i_k})^{\dagger}U^{(k)}_{i_k}=I$ of the original Hilbert space and
$U^{(k)}_{i_k}(U^{(k)}_{i_k})^{\dagger}=I$ of the support space of reduced density
matrix. While the total Hilbert space might change under
such an operation, the entanglement structure of the state
remains intact and hence the state remains in the same phase
with the same topological order.

If the class of systems under consideration have further symmetry
constraints, two Hamiltonians are in the same phase if they can be connected by a smooth path that stays within this symmetric region. Correspondingly, the equivalence relation between gapped ground states of the same phase needs to be
modified: \cite{CGW1035} \textit{If the class of systems have certain
symmetry, two gapped ground states belong to the same phase
if and only if they are related by a LU
transformation which does not break the symmetry.} Such an
restricted equivalence relation leads to a phase diagram
with more structure, as shown in Fig. \ref{fig:phase}(b).
First not all short range entangled states belong to the
same phase.  short range entangled states with different
symmetry breaking will belong to different phases. Those
symmetry breaking phases with short range entanglement
(SB-SRE) are well described by Landau's symmetry breaking
theory \cite{L3726,GL5064}.

Can all phases with short range entanglement be described by
symmetry breaking? Landau's symmetry breaking theory suggest
that states with the same symmetry always belong to the same
phase, which implies that all phases with short range
entanglement are described by symmetry breaking. However,
this result turns out to be not quite correct.  States
which do not break any symmetry can still belong to
different phases as well.\cite{Wqoslpub} We will refer to
the order in symmetric short range entanglement (SY-SRE)
states as \textit{symmetry protected topological
order}\cite{GW0931}.  For example, in the presence of
parity symmetry, the Haldane phase and the $S^z=0$ phase of
spin-1 chain belong to two different phases even though
both phases have short range entanglement and do not break the parity symmetry.  Also, in the
presence of time-reversal symmetry, the topological
insulators and the band insulators belong to two different
phases. Again both phases have short range entanglement.

For systems with long range entanglement, the phase diagram
similarly divides into symmetry breaking (SB-LRE) and
symmetric (SY-LRE) phases. The charge $4e$ superconducting states\cite{KLW0902}
and the symmetric $\Z_2$
states\cite{Wqoslpub,KW0906}are examples of the
SB-LRE phases and the SY-LRE phases respectively.

\section{1D Gapped Spin Systems and Matrix Product States}
\label{1DGapMPS}

Having defined the universality classes of phases as the
equivalence classes of states under (symmetric) local
unitary transformations, we would then like to know which
phases exist, or in other words, to classify all possible
phases in strongly correlated systems. Some partial
classifications have been discussed for strongly correlated
systems through string-net states,\cite{LWstrnet} and for
free fermion systems with certain symmetries through
K-theory.\cite{K0986,RSF0957} In this paper, we would like to consider 1D gapped strongly
correlated spin systems both with and without symmetry, and try to classify all
such systems whose ground state does not break any symmetry. (In other
words, the ground state has the same symmetry as the
Hamiltonian.)

Completely classifying strongly correlated spin systems seems to be a
hard task as in general strongly interacting quantum
many-body systems are very hard to solve. However, the
recent insight about describing 1D gapped ground states of spin systems with
matrix product state formalism\cite{FNW9243,PVW0701} provides us
with a handle to deal with this problem. A matrix product
states (MPS) is expressed as
\begin{equation}
\label{MPS}
|\phi\rangle = \sum_{i_1,i_2,...,i_N}
\Tr (A^{[1]}_{i_1}A^{[2]}_{i_2}...A^{[N]}_{i_N})|i_1i_2...i_N\rangle
\end{equation}
where $i_k=1...d$ with $d$ being the physical dimension of a
spin at each site, $A^{[k]}_{i_k}$'s are $D\times D$
matrices on site $k$ with $D$ being the inner dimension of
the MPS. It has been shown that matrix product states
capture the essential features of 1D gapped ground state,
for example an entanglement area law\cite{H0724} and
finite correlation length\cite{H0402,HK0681}, and provide an
efficient description of such states\cite{SWV0804}. On the
other hand, generic matrix product states satisfying a
condition called `injectivity' are all gapped ground states
of local Hamiltonians\cite{FNW9243,PVW0701}. Therefore, studying
this class of MPS will enable us to give a full
classification of 1D gapped spin systems.

Now the question of what phases exist in 1D gapped spin
systems can be restated as what equivalence classes of
matrix product states exist under LU
transformations. \Ref{VCL0501} gave a specific way
to apply such  LU transformations, which realizes
a renormalization group transformation on MPS that removes
local entanglement and takes the states to a simple fixed
point form. A partial classification of MPS is also given in
\Ref{VCL0501}. In the following we will use this
procedure to classify gapped phases of 1D spin system, in
particular the 1D systems with various symmetries.  We see
that the possible phases in 1D strongly correlated systems depend on
the symmetry of the class of systems.

First we will briefly review how the renormalization group
transformation \cite{VCL0501} is done. For the identification and optimal
removal of local entanglement, a particularly useful
mathematical construction is the double tensor
$\mathbb{E}^{[k]}_{\alpha\gamma,\beta\chi}=\sum_{i}
A^{[k]}_{i,\alpha\beta} \times (A^{[k]}_{i,\gamma\chi})^*$ of the MPS.
$\mathbb{E}^{[k]}$ uniquely determines the state up to a local
change of basis on each site\cite{NieC00,PVW0701}, that is, if
\begin{align}
 \mathbb{E}^{[k]}_{\alpha\gamma,\beta\chi}=\sum_{i}
A^{[k]}_{i,\alpha\beta} \times (A^{[k]}_{i,\gamma\chi})^*
=
\sum_{i}
B^{[k]}_{i,\alpha\beta} \times (B^{[k]}_{i,\gamma\chi})^*
\end{align}
then $A^{[k]}_{i,\alpha\beta}$ and $B^{[k]}_{i,\alpha\beta}$ are related
by a unitary transformation $U^{[k]}$:
\begin{align}
B^{[k]}_{i,\alpha\beta}= \sum_j U^{[k]}_{ij} A^{[k]}_{j,\alpha\beta} .
\end{align}
Therefore, states described by $A^{[k]}_i$ and $B^{[k]}_i$
have exactly the same entanglement structure, which is
faithfully captured in $\mathbb{E}^{[k]}$. A proof of this
fact can be found in \Ref{NieC00}. For
clarity, we present the proof in appendix \ref{dbt} following the notation of this paper.

Local unitary operations on MPS can be applied through manipulation of $\mathbb{E}^{[k]}$. Treat $\mathbb{E}^{[k]}$ as
a $D^2\times D^2$ matrix with row index $\alpha\gamma$ and
column index $\beta\chi$. To apply a unitary operation on $n$ consecutive sites, we combine the double tensor of the $n$ sites together into
\begin{align}
\label{tEE}
\tilde{\mathbb{E}}
=\mathbb{E}^{[1]} \mathbb{E}^{[2]}...  \mathbb{E}^{[n]}
\end{align}
and then decompose $\tilde{\mathbb{E}}$ into a
set of matrices $\t A_{\t i}$'s
\begin{align}
\t {\mathbb{E}}_{\alpha\gamma,\beta\chi}=\sum_{\t i}
\t A_{\t i,\alpha\beta} \times \t A_{\t i,\gamma\chi}^* .
\end{align}
Note $\t A_{\t i,\alpha\beta}$ is determined up to a
unitary transformation on $\t i$.  The index $\t i$ of $\t A_{\t
i,\alpha\beta}$, up to an unitary transformation, can be
viewed as the combination of $i_1,i_2,...,i_n$, the indices
of $A^{[1]}_{i_1,\alpha\beta}$, $A^{[2]}_{i_2,\alpha\beta}$,
...  $A^{[n]}_{i_n,\alpha\beta}$.  Going from original
indices $i_1,i_2,...,i_n$ to the effective index $\t i$
corresponds to applying a unitary operation on the
$n$-block and $\t A_{\t i}$ describes the new state after operation.

The unitary operation can be chosen so that local entanglement is maximally removed.
$\tilde{\mathbb{E}}$ contains all the information about the
entanglement of the block with the rest of the system
but not any detail of entanglement structure within the
block. Hence we can determine from $\tilde{\mathbb{E}}$ the optimal way of decomposition into $\t A$ which corresponds to the unitary operation that maximally removes local entanglement while
preserving the global structure. To do so, think of
$\tilde{\mathbb{E}}_{\alpha\gamma,\beta\chi}$ as a matrix
with row index $\alpha\beta$ and column index
$\gamma\chi$. It is easy to see that with such a
recombination, $\tilde{\mathbb{E}}$ is a positive matrix and
can be diagonalized
\begin{align}
 \tilde{\mathbb{E}}_{\alpha\gamma,\beta\chi}
=\sum_{\t i} \la_{\t i} V_{\t i,\al\bt} V^*_{\t i,\ga\del} ,
\end{align}
where we have kept only the non-zero eigenvalues $\la_{\t
i}$ and the corresponding eigenvectors $V_{\t i,\al\bt}$.
$\t A$ is then given by
\begin{align}
 \t A_{\t i,\alpha\beta}=\sqrt{\la_{\v i}} V_{\t i,\al\bt},
\end{align}
which are the matrices representing the new state. In retaining only the non-zero eigenvalues, we have reduced the physical dimension within the block to only those necessary for describing the entanglement between this block and the rest of the system. Local entanglement within the block has been optimally removed.

Each renormalization step in the renormalization procedure hence works by grouping every $n$
consecutive sites together and then applying the above
transformation to map $A^{[1]}$, $A^{[2]}$,..., $A^{[n]}$ to $\t
A$. So one renormalization step maps the original matrices
$(A^{[k]}_{i_k})^{(0)}$ on each site to renormalized
matrices $(A^{[k]}_{i_k})^{(1)}$ on each block.  Repeating
this procedure for a finite number of times corresponds to
applying a finite depth quantum circuit to the original
state. If the matrices reaches a simple fixed point form
$(A^{[k]}_{i_k})^{(\infty)}$ (up to local unitaries), we can
determine from it the universal properties of the phase
which the original state belongs to.

Such a renormalization procedure hence provides a way to
classify matrix product states under LU transformations by
studying the fixed point $(A^{[k]}_{i_k})^{(\infty)}$ that a
state flows to. Two states are within the same phase if and
only their corresponding fixed points states can be
transformed into each other by (symmetric) LU
transformations. In the following we will apply this method
to study short-range correlated matrix product states which
faithfully represent the class of 1D gapped ground states.

The short-range correlation is an extra constraint on the
set of matrix product states that we will consider. Not all
matrix product states describe gapped ground states of 1D
spin systems. In particular, 1D gapped ground states all
have finite correlation length\cite{HK0681} for equal time
correlators of any local operator, while matrix product
states can be long-range correlated. The finite correlation
length puts an extra constraint on MPS that the
eigen-spectrum of $\mathbb{E}$ should have a non degenerate
largest eigenvalue (set to be 1) (see appendix
\ref{ndg}).\cite{FNW9243,PVW0701} Therefore, we will assume
this property of $\mathbb{E}$ in our following discussion
and corresponding MPS will be called short-range
correlated (SRC) MPS.

This renormalization method is well
suited for the study of systems without translational invariance,
which we will discuss in detail. We will also make an
attempt to study translational invariant systems with this method. While the
full translational symmetry is reduced to block
translational symmetry in the RG process, by studying the
resulting equivalence classes for different values of block
size $n$, we expect to obtain a more complete classification
of translational invariant 1D systems. Indeed, the classification result is further confirmed by using a translational invariant LU transformation in the time evolution form to study equivalence between TI systems.

In the following sections, we will present our analysis and
results for different cases. First we will consider the situation where no specific
symmetry is required for the system.

\section{No topological order in 1D}
\label{notop}
When no symmetry is required for the class of system, we want to know what
kind of long range entanglement exists and hence classify
topological orders in 1D gapped spin systems. We will show
that
\[\frm{\emph{
All gapped 1D spin systems belong to the same phase
if there is no symmetry.
}}
\]
In other words, there is no topological order in 1D. This is similar to the generic case discussed in \Ref{VCL0501}.

To obtain such a result, we use the fact that gapped 1D spin
states\footnote{A state is a gapped state if there exist
a Hamiltonian $H$ such that the state is the non-degenerate
gapped ground state of $H$.} are described by short-range correlated (SRC) matrix product
states. Then one can show that all SRC
matrix product states can be mapped to product states with
LU transformations and hence there is no
topological order in 1D.

Consider a generic system without any symmetry (including translation symmetry) whose gapped ground state is described as an MPS with matrices $A^{[k]}_{i}$ that vary from site to
site. \Ref{PVW0701} gives a `canonical form' for the matrices so that the double tensor $\mathbb{E}^{[k]}_{\alpha\gamma,\beta\chi}$, when treated as a matrix with row index $\alpha\gamma$ and column index $\beta\chi$, has a left eigenvector $\Lambda^{[k]}_{\alpha\gamma}=\lambda^{[k]}_{\alpha}\delta_{\alpha\gamma}$ and corresponding right eigenvector $\Lambda^{[k+1]}_{\beta\chi}=\lambda^{[k+1]}_{\beta}\delta_{\beta\chi}$. Here $\lambda$'s are positive numbers and $\sum_{\alpha} \lambda^2_{\alpha}=1$. $\delta_{\alpha\gamma}=1$ when $\alpha=\gamma$ and $\delta_{\alpha\gamma}=0$ otherwise.\footnote{The convention chosen here is different from \Ref{PVW0701}, but equivalent up to an invertible transformation on the matrices $A^{[k]}_{i}$.} This eigenspace has the largest eigenvalue in $\mathbb{E}^{[k]}$\cite{EH7845} and is usually set to be $1$. Note that the right eigenvector on site $k$ is the same as the left eigenvector on site $k+1$ and has norm $1$, therefore when multiplying the double tensors together, this one dimensional eigenspace will always be of eigenvalue $1$.

There could be other eigenvectors of eigenvalue $1$ in $\mathbb{E}^{[k]}$. However,
this will lead to an infinite correlation
length\cite{FNW9243,PVW0701} and hence not possible in 1D
gapped state. Therefore, for short-range correlated MPS,
$\mathbb{E}^{[k]}$ must have a non-degenerate largest
eigenvalue $1$. When multiplying the double tensors
together, the remaining block of $\mathbb{E}^{[k]}$ will
decay exponentially with the number of sites. This
consideration is essential for determining the fixed point
of the renormalization procedure when applied to the MPS, as
shown below.

Now we apply the renormalization procedure as discussed in the previous section to remove local entanglement from a general SRC MPS. Take block size $n$. The double tensor on the renormalized sites are
given by $(\mathbb{E}^{[K]})^{(1)}=\prod_{k\in K}
(\mathbb{E}^{[k]})^{(0)}$, where $k$'s are the $n$ sites in
block $K$. ($\mathbb{E}$, again, is treated as a $D^2\times D^2$
matrix with row index $\alpha\gamma$ and column index
$\beta\chi$.)

After repeating the renormalization process a finite number
of times, $(\mathbb{E}^{[k]})^{(R)}$ will be arbitrarily
close to a fixed point form $(\mathbb{E}^{[k]})^{(\infty)}$
with non-degenerate eigenvalue $1$ and
$(\mathbb{E}^{[k]})^{(\infty)}_{\alpha\gamma,\beta\chi}=\t \Lambda^{[k]}_{\alpha\gamma}\t \Lambda^{[k+1]}_{\beta\chi}$,  where $\t \Lambda^{[K]}_{\alpha\gamma}=\t \lambda^{[k]}_{\alpha}\delta_{\alpha\gamma}$ and $\t \Lambda^{[k+1]}_{\beta\chi}=\t \lambda^{[k+1]}_{\beta}\delta_{\beta\chi}$.

\begin{figure}
\begin{center}
\includegraphics[width=3.0in]{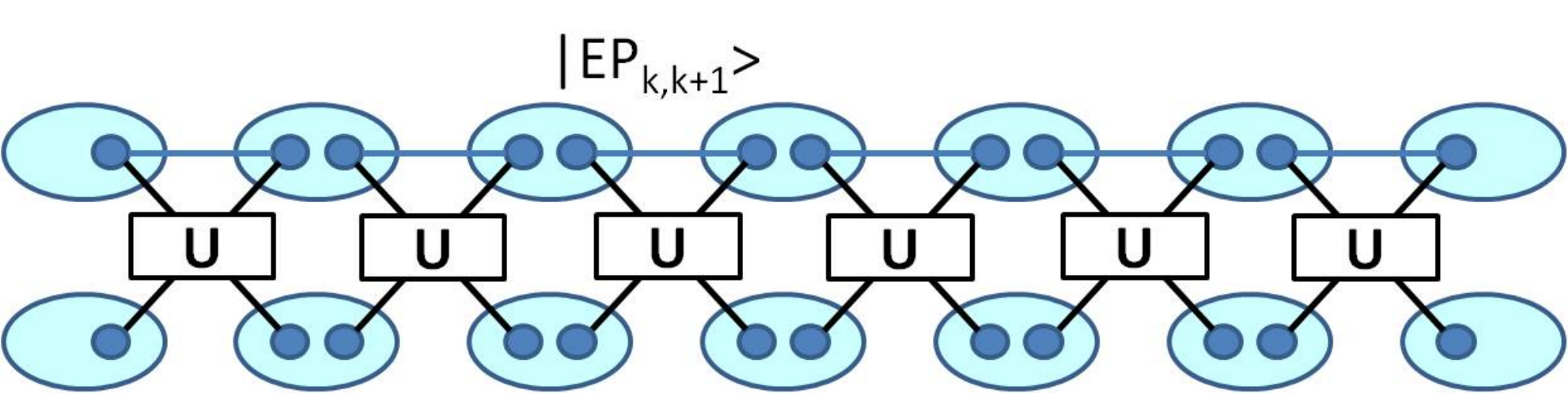}
\end{center}
\caption{(Color online) Disentangling fixed point state (upper layer,
product of entangled pairs) into direct product state (lower
layer) with LU transformations.}
\label{fig:FP}
\end{figure}

Now we can decompose $(\mathbb{E}^{[k]})^{(\infty)}$ into matrices to find the fixed point state.
One set of matrices giving rise to this double tensor is given by
\begin{align}
\label{Avv}
(A^{[k]}_{i^li^r,\alpha\beta})^{(\infty)}=\sqrt{\t \lambda^{[k]}_{i^l}} \delta_{i^l \alpha} \cdot \sqrt{\t \lambda^{[k+1]}_{i^r}} \delta_{i^r \beta}
\end{align}
$i^l,i^r=1...D$.
Here we use a pair of indices $(i^l,i^r)$ to label
the effective physical degrees of freedom on the
renormalized site $k$, and $(A^{[k]}_{i^l,i^r})^{(\infty)}$
is a set of matrices that defines the fixed-point MPS.
It is clear from the form
of the matrices that at fixed point every site is composed
of two virtual spins of dimension $D$. Every virtual spin is
in an entangled pair with another virtual spin on the
neighboring site $|EP_{k,k+1}\rangle =
\sum^D_{i=1} \t\lambda^{[k+1]}_{i}|i,i\rangle$ and the full many-body state
is a product of these pairs. An illustration of this state
is given in Fig. \ref{fig:FP}(upper layer).

Obviously we can further
disentangle these pairs by applying one layer of local
unitary transformations between every neighboring sites and
map the state to a product state (Fig. \ref{fig:FP}, lower layer).

Therefore, through these steps we have shown that all SRC
matrix product states can be mapped to product
states with LU transformations and hence there is
no topological order in 1D NTI system.

\section{Symmetry Protected Topological Order in 1D}
\label{symtop}

If the class of systems under consideration has certain
symmetry, the equivalence classes of states are defined in
terms of LU transformations that do not break the symmetry.
Therefore, when applying the renormalization procedure, we
should carefully keep track of the symmetry and make sure
that the resulting state has the same symmetry at each step.
Due to such a constrain on local unitary equivalence, we
will see that gapped ground states which do not break the
symmetry of the system divide into different universality
classes corresponding to different symmetry protected
topological orders. We will first discuss the case of
on-site symmetries in detail for non-translational invariant
(NTI) systems, i.e., the system has only an on-site symmetry
and no translation symmetry. Then we shall make an attempt
to study translational invariant (TI) systems, with the
possibility of having on-site symmetry or parity symmetry.
Lastly, we shall consider the case of time reversal symmetry.

\subsection{On-site Symmetry}
\label{usymm}
A large class of systems are invariant under on-site
symmetry transformations. For example, the Ising model is symmetric under
the $Z_2$ spin flip transformation and the Heisenberg model is
symmetric under $SO(3)$ spin rotation transformations. In this section, we will
consider the general case where the system is symmetric
under $u^{(0)}(g) \otimes...\otimes u^{(0)}(g)$ with
$u^{(0)}(g)$ being a unitary representation of a symmetry
group $G$ on each site. The representation can be linear or projective. That is, for any $g_1,g_2 \in G$,
\begin{equation}
u(g_1)u(g_2)=e^{i\theta(g_1,g_2)}u(g_1g_2)
\end{equation}
where $\theta(g_1,g_2)=0$ in a linear representation and $\theta(g_1,g_2)$ could take non-trivial value in a projective representation. A projective representation of a symmetry group is generally allowed in a quantum description of system because the factor $e^{i\theta(g_1,g_2)}$ only changes the global phase of a quantum state but not any physically measurable quantity. Therefore, in our classification, we will consider not only the case of linear representation, but also in general projective representations.

The on-site symmetry is the only symmetry required for the class of system. In particular, we do not require translational symmetry for the systems. However, for a simple definition of phase, we will assume certain uniformness in the state, which we will define explicitly in the following. We will classify possible phases for
different $G$ when the ground state is invariant (up to a
total phase) under such
on-site symmetry operations and is gapped (\ie short-range
correlated). Specifically, the ground state $|\phi_L\rangle$
on $L$ sites satisfies
\begin{align}
\label{u0act}
u^{(0)}(g) \otimes...\otimes
u^{(0)}(g) |\phi_L\rangle= \alpha_L(g)
|\phi_L\rangle
\end{align}
where $|\alpha_L(g)|=1$ are $g$ and $L$ dependent phase factors.

\subsubsection{On-site Linear Symmetry}
\label{usymm_l}

First, let us consider the simpler case where $u^{(0)}(g)$ form a linear representation of $G$. $\alpha_L(g)$ is then a one-dimensional linear representation of $G$.
Now we will try to classify these symmetric ground states using symmetric LU transformations and we find that
\[\frm{\emph{
Consider 1D spin systems with ONLY an
on-site symmetry $G$ which is realized linearly,
all the gapped phases that do not
break the symmetry are classified by $H^2(G, \C)$, the
second cohomology group of $G$, if $H^2(G, \C)$ is finite
and $G$ has a finite number of 1D representations.
}}
\]
We will also discuss the case of $U(1)$ group which has an
infinite number of 1D representations.

We will again assume that all gapped states can be represented
as short range correlated matrix product states.
We will use the renormalization
flow used before\cite{VCL0501} to simplify the matrix
product states and use the fixed-point matrix product
states to characterize different equivalent classes of LU
transformations, as two symmetric states belong to the same class if and only if their corresponding fixed-point states can be mapped to each other with symmetric LU transformations.

In order to compare different equivalent classes under
\emph{symmetric} LU transformations, it is
important to keep track of the symmetry while doing
renormalization.

First, in the renormalization procedure we group $n$ sites
together into a new site. The on-site symmetry
transformation becomes $\tilde{u}^{(0)}(g)=(\otimes
u^{(0)}(g))^n$, which is again a linear representation of
$G$.  The next step in RG transformation applies a unitary
transformation $w^{[k]}_1$ to the support space of new site
$k$. This is actually itself composed of two steps. First we
project onto the support space of the new site, which is the
combination of $n$ sites in the original chain. This is an
allowed operation compatible with symmetry $G$ as the
reduced density matrix $\rho_{n}$ is invariant under
$\tilde{u}^{(0)}(g)$, so the support space form a linear
representation for $G$. The projection of
$\tilde{u}^{(0)}(g)$ onto the support space
$P_n\tilde{u}^{(0)}(g)P_n$ hence remains a linear
representation of $G$. In the next step, we do some unitary
transformation $w^{[k]}_1$ within this support space which
relabels different states in the space. The symmetry
property of the state should not change under this
relabeling. In order to keep track of the symmetry of the
state, the symmetry operation needs to be redefined as
$(u^{[k]})^{(1)}(g)=w^{[k]}_1P_n\tilde{u}^{(0)}(g)P_n(w^{[k]}_1)^{\dagger}$.
After this redefinition, the symmetry operations
$(u^{[k]})^{(1)}(g)$ on each new site $k$ form a new
linear representation of $G$.

By redefining $(u^{[k]})^{(i)}(g)$ at
each step of RG transformation, we keep track of the
symmetry of the system.
Finally at the fixed point (\ie at a large RG step $i=R$), we
obtain a state described by
$(A^{[k]}_{i^li^r})^{(R)}$ which is
again given by the fixed point form \eqn{Avv}.
To describe a state that does not break the on-site
symmetry, here $(A^{[k]}_{i^li^r})^{(R)}$ is
invariant (up to a phase) under
$(u^{[k]})^{(R)}(g)$ on each site $k$.
Therefore\cite{GWS0802},
\begin{align}
\label{Eqn:LU_A}
 \sum_{j^lj^r} u^{[k]}_{i^li^r,j^lj^r}(g) A^{[k]}_{j^lj^r}
& =
\al_{[k]}^{(R)}(g)
N_{[k]}^{-1}(g)A^{[k]}_{i^li^r} M_{[k]}(g)
\nonumber\\
 N_{[k]}(g) &= M_{[k-1]}(g)
\end{align}
must be satisfied with some invertible matrix $N_{[k]}(g)$ and
$M_{[k]}(g)$. Here $k$ labels the coarse grained sites
and we have dropped the RG step label $R$ (except
in $\al_{[k]}^{(R)}(g)$).
Each coarse grained site is a combination of
$n^R$ original lattice sites and
$\al_{[k]}^{(R)}(g)$ form a 1D (linear) representation of $G$.

Solving this equation we find
the following results (see appendix \ref{soluu}):\\
(a) $N_{[k]}(g)$ and $M_{[k]}(g)$ are projective
representations of $G$ (see \eqn{MNproj}). Projective
representations of $G$ belong to different classes which
form the second cohomology group $H^2(G, \C)$ of $G$. (For a
brief introduction on projective representation, see
appendix \ref{prorep}). $M_{[k]}(g)$ and $N_{[k]}(g)$
corresponds to the same element $\om$ in
$H^2(G, \C)$.\\
(b) The linear symmetry operation $u^{[k]}(g) $ must be of
the form $\al_{[k]}^{(R)}(g)u^{[k],l}(g)\otimes u^{[k],r}(g)$ where
$u^{[k],l}$ and $u^{[k],r}$ are projective representations
of $G$ and correspond to inverse elements $\om$ and $-\om$ in $H^2(G,\C)$ respectively. $\al_{[k]}^{(R)}(g)$ is a 1D (linear) representation of $G$.
$u^{[k],l}$ and $u^{[k],r}$ act on the two virtual spins
separately (see \eqn{uaMN}).

Therefore, the fixed point state is formed by entangled
pairs $|EP_{k,k+1}\rangle$ of virtual spins which are
invariant, up to a phase (due to the non-trivial $\al_{[k]}^{(R)}(g)$),
under linear
transformation $u^{[k],r}(g)\otimes u^{[k+1],l}(g)$.

Now we use
the uniformness of the state and simplify our
discussions. Specifically, we assume that $\al_{[k]}^{(R)}(g)$
does not depend on the site index $k$. Certainly,
$\al_{[k]}^{(R)}(g)$ does not depend on $k$ if the state has
the translation symmetry. If the 1D representations of $G$
are discrete, then for weak randomness that slightly break
the translation symmetry, $\al_{[k]}^{(R)}(g)$ still does
not depend on $k$. So we can drop the $k$ index and
consider $\al^{(R)}(g)$.

Does different $\al^{(R)}(g)$ label different symmetric phases? First,
the answer is no if the number of 1D
representations of $G$ is finite (as is the case for $Z_n$,
$SO(3)$, etc). Because we can always choose block size $n$ properly so
that $\alpha^{(R)}(g)=1$ and the difference between symmetric
states due to $\alpha^{(R)}(g)$ disappears. In the case of $U(1)$
group, there are infinitely many different 1D
representations $e^{im\theta}$, labeled by integer $m$. For
two states with positive $m_1$, $m_2$, we can always choose
block size $m_2n^R$, $m_1n^R$ respectively, so that the 1D
representations become the same. This is also true for
negative $m_1$, $m_2$. But if $m_1$, $m_2$ take different
sign (or one of them is $0$), the 1D representations will
always be different no matter what blocking scheme we use.
Therefore, $U(1)$ symmetric states divide into three classes
due to different 1D representations that can be labeled by
$\{+,0,-\}$. After these considerations, we will ignore the 1D representations
$\al^{(R)}(g)$ in the following discussion.

We find that the entangled
pairs $|EP_{k,k+1}\rangle$ of virtual spins in the fixed
point state are exactly invariant under linear
transformation $u^{[k],r}(g)\otimes u^{[k+1],l}(g)$.
The left virtual spin of each site
forms a projective representation of $G$ corresponding to
element $\om$ in $H^2(G,\C)$, while the right virtual spin
corresponds to element $-\om$. In appendix \ref{equiv}, we
will show that fixed point states with the same $\om$ can be
related by a symmetric LU transformation, while those with
different $\om$ cannot. Therefore, the phases of SRC MPS
that are invariant under linear on-site symmetry of group
$G$ are classified by the second cohomology group
$H^2(G,\C)$.
(When $G=U(1)$, further division of classes due
to different 1D representations of $G$ exist. The
equivalence classes are labeled by $\alpha \in \{+,0,-\}$
and $\om \in H^2(U(1),\C)$.)

\subsubsection{On-site Projective Symmetry}
\label{usymm_p}

Due to the basic assumption of quantum mechanics that the global phase of a quantum state will not have any effect on the physical properties of the system, it is necessary to consider not only the linear representation of symmetry operations on the system, but also the projective representations. For example, on a half-integer spin, rotation by $2\pi$ is represented as $-I$, minus the identity operator instead of $I$. Hence, the rotation symmetry $SO(3)$ is represented projectively on half integer spins. In order to cover situations like this, we discuss in this section systems with on-site projective symmetry of group $G$.

Again, we consider the case when the ground state does not
break the symmetry, i.e.
$u^{(0)}(g) \otimes...\otimes
u^{(0)}(g) |\phi_L\rangle= \alpha_L(g)
|\phi_L\rangle$,
where $u^{(0)}(g)$ form a projective
representation of group $G$ corresponding to class $\om$.
Assuming uniformness of the state, we require that $\om$ does not vary from site to site.

But this can be reduced to the previous linear case. As long
as $H^2(G,\C)$ is finite and $\om$ has a finite order $n$,
we can take block size $n$ so that after blocking, the
symmetry operation on the renormalized site
$\tilde{u}^{(0)}(g)=(\otimes u^{(0)}(g))^n$ corresponds to
$n\om = \om_0$ in $H^2(G,\C)$. Therefore, the state after one blocking step is symmetric under an on-site linear representation of group $G$ and all the reasoning in the previous section applies. We find that the classification with projective on-site symmetry is the same as linear on-site symmetry. That is
\[\frm{\emph{
Consider 1D spin systems with ONLY an
on-site symmetry $G$ which is realized projectively,
all the gapped phases that do
not break the symmetry are classified by $H^2(G, \C)$, the
second cohomology group of $G$, if $H^2(G, \C)$ is finite
and $G$ has a finite number of 1D representations.
}}
\]
The $U(1)$ group does not have a non-trivial projective representation and will not introduce any complication here.

\subsubsection{Examples}

Since $G=\Z_n$ has no non-trivial projective representations,
we find that
\[\frm{\emph{
All 1D gapped systems with only on-site $\Z_n$ symmetry belong to the
same phase.
}}
\]

For spin systems with only spin rotation symmetry, $G=SO(3)$.
$SO(3)$ has two types of projective representations
described by $H^2(SO(3),\C)=\{0,1\}$, corresponding to integer and half-integer spin representations.
We find that
\[\frm{\emph{
For integer-spin systems,
all 1D gapped systems with only on-site $SO(3)$ spin rotation
symmetry have two different phases.
}}
\]
Such a result has some relation to a well known
result\cite{HY9783} for NTI spin-1 Heisenberg chain
\begin{align}
\label{rspin}
 H=\sum_i J_i \v S_i\cdot \v S_{i+1}.
\end{align}
The model undergoes an impurity driven second order phase
transition from the Haldane phase\cite{H8364} to the random
singlet phase\cite{MDH7934,F9499} as the randomness in $J_i$
increases.

For half-integer-spin systems, $SO(3)$ is represented projectively on each site, yet the classification is the same as the integer case.
we find that
\[\frm{\emph{
For half-integer-spin systems,
all 1D gapped states with only on-site $SO(3)$ spin rotation
symmetry have two different phases.
}}
\]
Representative states of the two phases are nearest-neighbor
dimer states, but with dimer between sites $2i$ and $2i+1$
in the first phase and between sites $2i-1$ and $2i$ in the
second phase.

The projective representation of $SO(3)$ on
half-integer-spins form a linear representation of $SU(2)$.
If we think of the linear representation of $SO(3)$ on
integer-spins as a (unfaithful) linear representation of
$SU(2)$ and allow the mixture of integer and half-integer
spins on one site, then the two phases of $SO(3)$ merge into
one. Therefore, systems with only on-site $SU(2)$ symmetry
(which implies the mixture of integer and half-integer spins
on each site) belong to one phase as we can map integer-spin
singlets into half-integer-spin singlets without breaking
the $SU(2)$ symmetry (see appendix \ref{equiv}). Such a
procedure breaks down if $SO(3)$ symmetry is required for
each site as the direct sum of a linear representation (on
integer-spin) and a projective representation (on
half-integer-spin) is no long a projective representation
for $SO(3)$.

In this way, we have obtained a full classification of
the phases of gapped NTI 1D spin systems with various on-site
symmetry.

\subsection{Translation Invariance}
\label{TIsym}
We have discussed the gapped phases of 1D NTI systems
that have some on-site symmetries. In this section, we
would like to discuss translation symmetric systems. We
will consider those  translation symmetric systems whose
ground states are gapped and translation invariant (TI).\footnote{In general,
a state which does not break the translational symmetry of
the system is invariant under translation up to a phase.
That is, the state carries a finite momentum. However, we
will restrict ourselves only to the case where the ground
state has zero momentum and we will say the states are
translational invariant instead of translational symmetric.}

\subsubsection{$\{n_i\}$-block TI LU transformations and b-phases}

To discuss the TI phases, we need to discuss the equivalence
classes under TI LU transformations.  However, it is hard to
use quantum circuit to describe TI LU transformations.
Thus, the quantum circuit formulation used in this paper is
inconvenient to describe TI LU transformations. However, in
this section, we will first try to use the quantum circuit
formulation of LU transformations to discuss the phases of
TI gapped states. While we are able to identify many different phases in this way, we can not rigorously prove the equivalence of states within each phase using the non-translational invariant circuit. In order to establish this equivalence, we later apply the other formulation of local unitary transformation -- a finite time evolution with a local Hamiltonian(Eqn.\ref{LU_H}) where TI can be preserved exactly and we confirm the classification result obtained with quantum circuits(see appendix \ref{fullTI}).

\begin{figure}
\begin{center}
\includegraphics[scale=0.5]{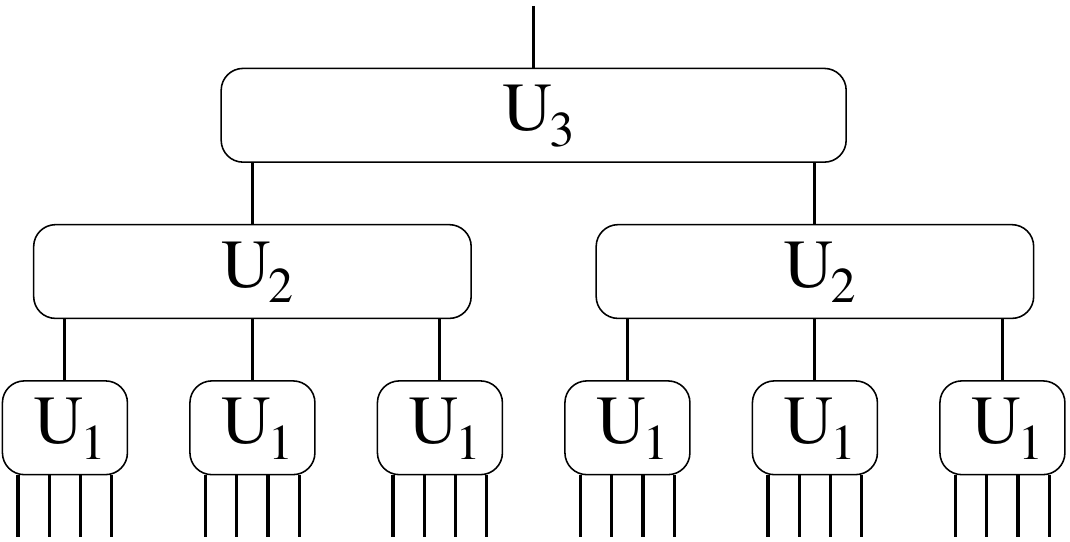}
\end{center}
\caption{
A $\{n_i\}$-block TI LU transformation
described by a quantum circuit of three layers. Here
$n_1=4$, $n_2=3$, $n_3=2$.  The unitary transformations
$U_i$ on different blocks in the $i^{th}$ layer are the
same.
}
\label{nbqc}
\end{figure}

Let us consider the LU transformations represented by
quantum circuits which is formed by the unitary operators on
blocks of $n_i$ sites in the $i^{th}$ layer (see Fig.
\ref{nbqc}).  We will call such LU
transformations $\{n_i\}$-block LU transformations.  If the
unitary operators on different blocks in the same layer of
the quantum circuit are the same, we will call the LU
transformations a $\{n_i\}$-block TI LU
transformations.

In this paper, we will try to use quantum circuit
formulation of the LU transformations to discuss gapped
translation symmetric phases.  One way to do so is to use
the equivalent classes of the $\{n_i\}$-block TI LU
transformations to classify the gapped translation symmetric
phases.  However, since the $\{n_i\}$-block TI LU
transformations are different from the TI LU
transformations, the equivalent classes of $\{n_i\}$-block
TI LU transformations are different from the equivalent
classes of the TI LU transformations.  But since the TI LU
transformations are special cases of $\{n_i\}$-block TI LU
transformations, each equivalent class of $\{n_i\}$-block TI
LU transformations are formed by one or more equivalent
classes of TI LU transformations.

Therefore, we can use the equivalent classes of
$\{n_i\}$-block TI LU transformations to describe the gapped
TI phases, since different equivalent classes of
$\{n_i\}$-block TI LU transformations
always represent different TI quantum phases. On the other
hand, the equivalent classes of $\{n_i\}$-block TI LU
transformations may not separate all gapped TI phases.
Sometimes, a single equivalent class of $\{n_i\}$-block TI LU
transformations may contain several different gapped TI
phases.

To increase the resolution of the $\{n_i\}$-block
TI LU transformations, we would like to
introduce the block equivalent classes: two states belong to
the same block equivalent classes, iff, \emph{for all values of $n_i$}, they can be mapped
into each other through $\{n_i\}$-block TI LU transformations.

Clearly, each block equivalent class might still contain
several different gapped TI phases. In this paper, we will
call a block equivalent class a block phase, or b-phase.
It is possible that the block equivalent classes are the
same as the universal classes that represent the gapped TI
phases. In this case the gapped TI b-phases are the same as
the gapped TI phases, and we can study the gapped TI phases
through block equivalent classes.
In the following, we will first study the b-phases for 1D strongly
correlated systems with translation and some other
symmetries and then confirm that gapped TI b-phases do coincide with gapped TI phases.

To describe TI gapped states we will use TI MPS representation with site independent matrices.
But how do we know a TI gapped state can always
be represented by a MPS with site independent matrices?
A non-uniform MPS in \eqn{MPS} can represent a TI state
if the matrices $A^{[k]}_i$ satisfy, for example,
\begin{align}
 A^{[k+1]}_i= M^{-1} A^{[k]}_i M
\end{align}
for an invertible matrix $M$.

It is proven in \Ref{PVW0701} that every TI state does have a TI MPS representation. Specifically, in the example considered above, we can transform the matrices so that they become site independent.
Let us introduce
\begin{align}
\t A^{[k]}_i= N_{[k-1]}^{-1} A^{[k]}_i N_{[k]}.
\end{align}
By construction, $A^{[k]}_i$ and $\t A^{[k]}_i$
will represent the same MPS.
We see that
\begin{align}
 \t A^{[k+1]}_i &= N_{[k]}^{-1} A^{[k+1]}_i N_{[k+1]}
\nonumber\\
&=
N_{[k]}^{-1}M^{-1} A^{[k]}_i M N_{[k+1]}
\\
&=
N_{[k]}^{-1}M^{-1} N_{[k-1]} \t A^{[k]}_i N_{[k]}^{-1} M N_{[k+1]}
\nonumber
\end{align}
We see that if we choose
\begin{align}
 N_{[k]}=M^{-k},
\end{align}
we will have
\begin{align}
 \t A^{[k+1]}_i=\t A^{[k]}_i.
\end{align}
Hence reducing the original non-uniform representation to a uniform one.

We also like to remark that by ``TI gapped state'', we mean
that there exists a TI gapped Hamiltonian $H_L$ on a lattice
of $L$ sites such that the TI gapped state is the ground
state of $H_L$. We would like to stress that we do not need the above
condition to be true for all values of $L$.  We only require
$H_L$ to be gapped for a sequence of lattice sizes $\{L_i\}$
that $\lim_{i\to\infty} L_i=\infty$. In this paper, when we
discuss a system of size $L$, we always assume that $L$
belongs to such a sequence $\{L_i\}$. This consideration is important if we want to include in our discussion, for example, boson systems with 1/2 particles per site which can only be defined for even system size $L$.

\subsubsection{Gapped TI b-phases coincide with gapped TI phases}

In the above discussion, we see that for a large block size,
the related matrix [see \eq{tEE}] $\tilde{\mathbb{E}}
=\mathbb{E}^{[1]} \mathbb{E}^{[2]}...  \mathbb{E}^{[n]}$ is
dominated by its largest eigenvalue.  For translation
invariant states, $\mathbb{E}^{[i]}=\mathbb{E}$ and
$\tilde{\mathbb{E}}=\mathbb{E}^n$.  So the fixed-point
$\tilde{\mathbb{E}}$, and hence the gapped TI b-phase,
is directly determined by the largest
eigenvalue and the corresponding left and right eigenvectors
of $\mathbb{E}$.

Since a gapped TI b-phase is directly determined
by $\mathbb{E}$ on each site, we do not need to do any blocking
transformation
to understand the gapped TI b-phase. We can directly extract
its fixed-point tensor from $\mathbb{E}$. This suggests that
gapped TI b-phases coincide with gapped TI phases.  Indeed,
we can directly deform $\mathbb{E}$ to the fixed-point
$\tilde{\mathbb{E}}$ by changing other non-largest eigenvalues
to zero.  Such a procedure allows us to deform the matrix
$A_{i,\al\bt}$ of the initial MPS to the fixed-point matrix
$\t A_{i,\al\bt}$ of the final MPS. Since the largest
eigenvalue of $\mathbb{E}$ is not degenerate, the state remains short range
correlated (and gapped) during the deformation.\cite{FNW9243,PVW0701} (As
we are writing up the above arguments, we learned that a
similar method is used by Schuch et.  al.\cite{SC10}) Also
the state does not break the translation symmetry (unlike
the $n$-block transformations) and other symmetries during
the deformation.  This allows us to show that the gapped TI
phases is characterized by the largest eigenvalue and the
corresponding left and right eigenvectors of $\mathbb{E}$.
Thus gapped TI b-phases coincide with gapped TI phases.  For
details, see appendix \ref{fullTI}.

In the following, we will use the
$\{n_i\}$-block TI LU transformations to discuss various
gapped TI phases of 1D systems.

\subsubsection{1D systems with only translation symmetry}
\[\frm{\emph{
For 1D systems with only translation symmetry,
there is only one gapped TI phase.
}}
\]
Such a result generalizes the earlier result that for 1D
systems with no symmetry, there is only one gapped phase.
To obtain the new result, we basically repeat what we did
in section \ref{notop}. The only difference is that the matrices
representing the state now are site independent and in
section \ref{notop} we use $\{n_i\}$-block LU
transformations to reduce the 1D NTI MPS, while to derive
the new result, here we use $\{n_i\}$-block TI LU
transformations to reduce the 1D TI MPS.

\subsubsection{1D systems with translation and on-site unitary
symmetries}

Similarly, by repeating the discussions in section \ref{usymm_l} and section \ref{usymm_p} for
$\{n_i\}$-block TI LU transformations on the 1D TI MPS, we
can show that
\[\frm{\emph{
For a 1D spin system with translation and an
on-site projective symmetry $u(g)$, the symmetric ground state
cannot be short-range correlated, if the projective
symmetry $u(g)$ correspond to a non-trivial elements in
$H^2(G,\C)$.
}}
\]
The reason is as follows. If a 1D state with translation
symmetry is short-range correlated, it can be
represented by a TI MPS.  Its fixed-point MPS also has an
on-site projective unitary symmetry $\t u(g)$.  For a proper
choice of block size $n$, we can make $u(g)$ and $\t u(g)$
to be the same type of projective representation described by
$\om_{sym}\in H^2(G,\C)$.  For TI fixed-point MPS, we have
$\om_{[k]}=\om_{[k-1]}$ since $M_{[k]}(g)=M_{[k-1]}(g)$(cf. appendix \ref{soluu}). Thus
$\om_{sym}=0$, that is, the trivial element in  $H^2(G,\C)$. So, if
$\om_{sym}\neq 0$, the 1D TI state cannot be short-range
correlated. In other words
\[\frm{\emph{
1D spin systems with translation and an
on-site projective symmetry
are always gapless or have degenerate ground states that
break the symmetries.
}}
\]
If the ground state of the 1D spin system does not break
the on-site symmetry and the translation symmetry, then
ground state is not short-range correlated and is gapless.  If
the ground state of the 1D spin system breaks the on-site
symmetry or the translation symmetry, then the ground state
is degenerate.

As an application of the above result, we find that
\[\frm{\emph{
1D half-integer-spin systems with translation and the $SO(3)$
spin rotation symmetry
are always gapless or have degenerate ground states.
}}
\]
which agrees with the well known result of \Ref{LSM6107}.

To have a gapped TI 1D state with an on-site symmetry, the
symmetry must act linearly (\ie not projectively).  In this
case,
we can show that the total phase factor of the state $\al_L(g)$ breaks up into $L$ 1D representations $\al(g)$ (see appendix \ref{alL})
\[\frm{\emph{
For 1D spin systems of $L$ sites with translation and
an on-site symmetry $G$, a gapped
state that do not break the two symmetries
must transform as
\begin{align}
\label{u0act_TI}
u^{(0)}(g) \otimes...\otimes
u^{(0)}(g) |\phi_L\rangle= [\alpha(g)]^L
|\phi_L\rangle
\end{align}
for all values of $L$ that is large enough.
}}
\nonumber
\]
Here $u^{(0)}(g)$ is the linear representation of $G$ acting
on the physical states in each site and $\al(g)$ is an
one-dimensional linear representation of $G$.

Let us apply the above result to a boson system with $p/q$
bosons per site.  Here the bosons number is conserved and
there is an $U(1)$ symmetry.  Certainly, the system is well
defined only when the number of sites $L$ has a form $L=J q$
(assuming $p$ and $q$ have no common factors).  For such
$L$, we find that $\al_L(g)=\al_0(g)^{J}=\al_0(g)^{L/q}$,
where $\al_0(g)$ is the generating 1D representation of the
$U(1)$ symmetry group.  So \eqn{u0act_TI} is \emph{not}
satisfied for some large $L$.  Therefore,
\[\frm{\emph{
a 1D state of
conserved bosons with fractional bosons per site must be
gapless, if the state does not break the $U(1)$ and the
translation symmetry.
}}
\]

In higher dimensions, the situation
is very different.  A 2D state of conserved bosons with
fractional bosons per site can be gapped, and, at same time,
does not break the $U(1)$ and the translation symmetry.
2D fractional quantum Hall states of bosons
on lattice provide examples for such kind of states.

Also, by repeating the discussions in section \ref{usymm_l}
for $\{n_i\}$-block TI LU transformations on the 1D TI MPS,
we can show that
\[\frm{\emph{
For 1D spin systems with only translation and
an on-site linear symmetry $G$, all the phases of gapped
states that do not break the two symmetries are classified
by a pair $(\om,\al)$ where $\om\in H^2(G,\C)$ label
different types of projective representations of $G$ and
$\al$ label different 1D representations of $G$.
}}
\]
Here $\alpha(g)$ is an 1D representation of $G$
that appear in \eqn{u0act_TI}.
The symmetric LU transformations cannot change 1D representation
$\al(g)$.  So the different phases are also distinguished
by the 1D representations $\al$ of $G$.

Here are a few concrete examples:
If we choose the symmetry group to be $G=\Z_n$, we find
\[\frm{\emph{
For 1D spin systems with only translation and
on-site $\Z_n$ symmetry, there are $n$ phases for gapped
states that do not break the two symmetries.
}}
\]
This is because $\Z_n$ has no projective representations
and has $n$ different 1D representations.
As an example, consider the following model
\begin{align}
 H=\sum_i [ - h \si^z_i - \si^x_{i-1} \si^y_i \si^z_{i+1} ] ,
\end{align}
where $\si^{x,y,z}$ are the Pauli matrices.  The model has a
$Z_2$ symmetry generated by $\si^z$.  The two
different $Z_2$ symmetric phases correspond the $h\to
\infty$ phase and the $h\to -\infty$ phase of the model.

If we choose the symmetry group to be $G=SO(3)$, we find
\[\frm{\emph{
For 1D integer-spin systems with only translation and
$SO(3)$ spin rotation symmetry, there are two phases for gapped
states that do not break the two symmetries.
}}
\]
This is because $SO(3)$ has only one 1D representation and
$H^2(SO(3),\C)=\Z_2$.  Such a result agrees with the well
known result that the AKLT state\cite{AKL8799} of spin-1
chain and the direct product state with spin-0 on each site
represent two different $SO(3)$ symmetric TI phases.  The
AKLT state (and the related Haldane phase\cite{H8364}) has
gapless boundary spin-1/2 states\cite{HKA9081,GGL9114,N9455}
and non-trivial string orders,\cite{NR8909,KT9204} which
indicate that the AKLT state is really different from the
spin-0 product state. Actually, the full symmetry of $SO(3)$ can be relaxed to only the dihedral group $D_2$($\Z_2 \times \Z_2$) of rotation by $\pi$ around $x$, $y$ and $z$ axis. As explained in appendix \ref{prorep}, $D_2$ has one non-trivial projective representation, to which the AKLT state corresponds. AKLT is different from the spin-0 product state as long as on-site $D_2$ symmetry is preserved. This is consistent with the result in \Ref{KT9204,PBT0959}.

\subsubsection{1D systems with translation and parity symmetries}

In this section, we will consider the case of parity
symmetry for translational invariant system. We define the parity
operation $P$ for a spin chain to be in general composed of two
parts: $P_1$, exchange of sites $n$ and $-n$; $P_2$, on-site
unitary operation $u^{(0)}$ where $(u^{(0)})^2=I$.
\footnote{The $Z_2$ operation $u^{(0)}$ is necessary in the definition of parity if we want to consider for example, fixed point state with $|EP\rangle=|00\rangle+|11\rangle$ be to parity symmetric. The state is not invariant after exchange of sites, and only maps back to itself if in addition the two virtual spins on each site are also exchanged.}
Similar to the previous discussion, $P$ gets redefined as we
renormalize the state until at fixed point $P_1$ becomes the
exchange of renormalized sites and $P_2$ becomes
$u^{(\infty)}$ on every site, $(u^{(\infty)})^2=I$. The
fixed point matrices hence satisfies(the $\infty$ label is dropped):
\begin{equation}
\sum_{j^lj^r} u_{i^li^r,j^lj^r} A_{j^lj^r}^T =\pm M^{-1}A_{i^li^r}M
\label{PA_A}
\end{equation}
for some invertible matrix $M$, where we have used that the
1D representation of parity is either $(1,1)$ or $(1,-1)$. We label the two 1D representations with $\alpha(P)=\pm 1$.
Here $M$ satisfies $M^{-1}M^T=e^{i\theta}$. But $M=(M^T)^T=e^{2i\theta}M$, therefore, $e^{i\theta}=\pm 1$ and correspondingly $M$ is either symmetric $M=M^T$ or antisymmetric $M=-M^T$. We will label this sign factor as $\beta(P)=\pm 1$.

Solving this equation gives that $u=\alpha(P) v(u^l\otimes u^r)$,
where $v$ is the exchange operation of two virtual spins
$i^l$ and $i^r$ and $u^l$,$u^r$ act on $i^l$,$i^r$
respectively. $(u^l)^T=\beta(P)u^l$ and $(u^r)^T=\beta(P)u^r$.
It can then be shown that each entangled pair
$|EP_{k,k+1}\rangle$ must be symmetric under parity
operations and satisfies $u^r_k\otimes
u^l_{k+1}|EP_{k+1,k}\rangle = \alpha(P) |EP_{k,k+1}\rangle$. There are hence four different symmetric phases corresponding to $\alpha(P)=\pm 1$ and $\beta(P)=\pm 1$.
By enlarging the local Hilbert space, we can show similarly
as before that fixed points within each class can be mapped
from one to the other with the $\{n_i\}$-block TI LU
transformation preserving the parity symmetry. On the other hand, fixed points in different classes can not be connected without breaking the symmetries. Therefore,
under the $\{n_i\}$-block TI LU transformation, there are
four block classes with parity symmetry and hence four
parity symmetric TI phases:
\[\frm{\emph{
For 1D spin systems with only translation and
parity symmetry, there are four phases for gapped
states that do not break the two symmetries.
}}
\]

As an example, consider the following model
\begin{align}
 H=\sum_i [ -B S^z_i + \v S_i\cdot \v S_{i+1} ] ,
\end{align}
where $\v S_i$ are the spin-1 operators.  The model has a
parity symmetry. The $B=0 $ phase and the $B\to +\infty$
phase of the model correspond to two of the four phases discussed above.
The $B=0$ state\cite{H8364} is in the
same phase as the AKLT state. In the fixed-point state for
such a phase, $|EP_{k,k+1}\rangle = |\up\down\>-
|\down\up\>$. The parity transformation exchange the first
and the second spin, and induces a minus sign: $P:
|EP_{k,k+1}\rangle \to -|EP_{k,k+1}\rangle $.  The $B\to
+\infty$ state is the $S^z=1$ state.  Its entangled pairs are
$|EP_{k,k+1}\rangle = |\up\up\>$ which do not change sign
under the parity transformation.  Thus the stability of the
Haldane/AKLT state is also protected by the parity
symmetry.\cite{BTG0819,GW0931,PBT0959}

\begin{figure}
\begin{center}
\includegraphics[width=3.0in]{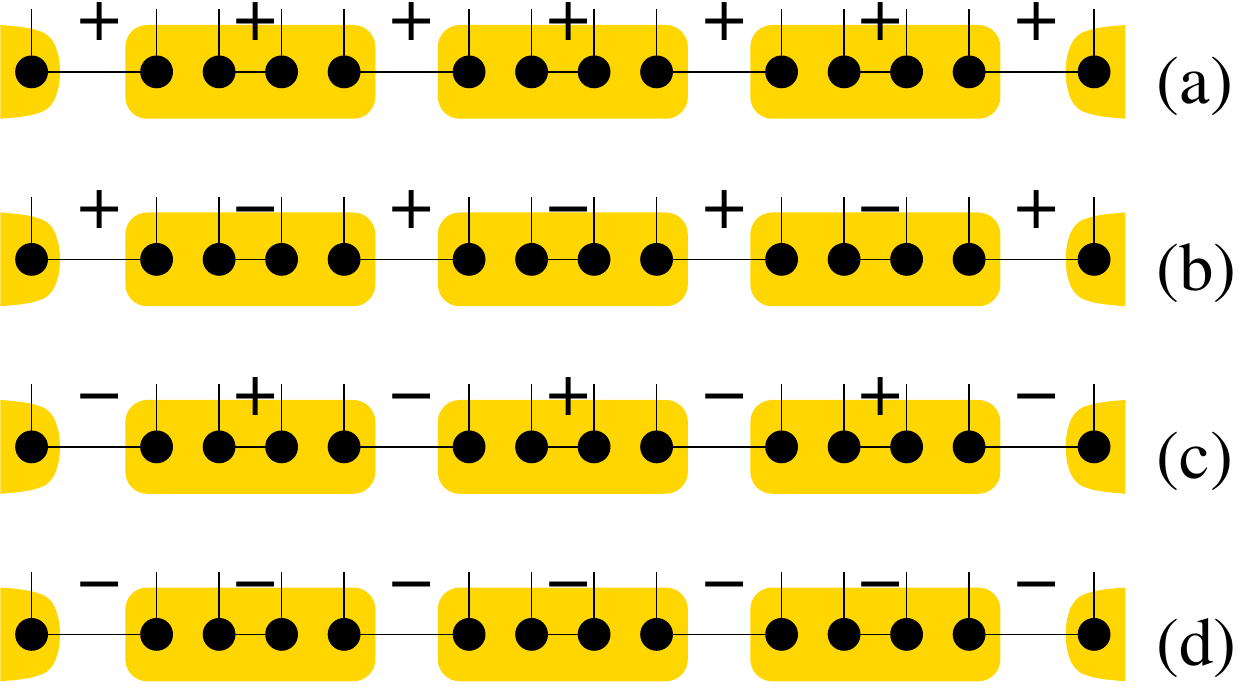}
\end{center}
\caption{
(Color online)
Representative states of the four parity symmetric phases, each corresponding to (a) $\alpha(P)=1$, $\beta(P)=1$ (b) $\alpha(P)=-1$, $\beta(P)=1$ (c) $\alpha(P)=-1$, $\beta(P)=-1$ (d) $\alpha(P)=1$, $\beta(P)=-1$. $+$ stands for a parity even entangled pair (e.g. $|00\rangle+|11\rangle$), $-$ stands for a parity odd entangled pair (e.g. $|01\rangle-|10\rangle$). Each site contains four virtual spins.}
\label{fig:4Pstates}
\end{figure}

To understand why there are four parity symmetric phases instead of two (parity even/parity odd), we give four representative states in Fig. \ref{fig:4Pstates}, one for each phase. Connected pair of black dots denotes an entangled pair. $+$ stands for a parity even pair, for example $|00\rangle+|11\rangle$, and $-$ stands for a parity odd pair, for example $|01\rangle-|10\rangle$. Each rectangle corresponds to one site, with four virtual spins on each site. The four states are all translational invariant. If the parity operation is defined to be exchange of sites together with exchange of virtual spins $1$ and $4$, $2$ and $3$ on each site, then states (a) and (d) are parity even while (b) and (c) are parity odd. But (a) and (d) (or (b) and (c)) are different parity even (odd) states and cannot be mapped to each other through local unitary transformations without breaking parity symmetry. Written in the matrix product representation, the matrices of the four states will transform with $\alpha(P)=\pm 1$ and $\beta(P)=\pm 1$ respectively. Therefore, the parity even/odd phase breaks into two smaller phases and there are in all four phases for parity symmetric systems.

\subsection{Time Reversal Symmetry}
\label{TR}
Time reversal, unlike other symmetries, is represented by antiunitary operator $T$, which is equivalent to the complex conjugate operator $K$ followed by a unitary operator $U$. $T$ has two projective representations: one on integer spins with $T^2=I$ and the other on half-integer spins with $T^2=-I$. The classification of gapped 1D time reversal invariant phases follows closely the cases discusses before. In this section, we will highlight the differences and give our conclusion.

First, a state $|\phi\rangle$ is called time reversal invariant if $T\otimes T ...\otimes T |\phi\rangle = \bt |\phi\rangle$, where $|\bt|=1$. But for anti-unitary $T$, the global phase $\bt$ is arbitrary and in particular we can redefine $|\phi'\rangle=\sqrt{\bt}|\phi\rangle$, such that $T\otimes T ...\otimes T |\phi'\rangle = |\phi'\rangle$. Therefore, in the following discussion, we will assume WLOG that $\bt=1$.

Now let us consider system without translational invariance.
$T^2=I$ or $-I$ does not make a difference here as we can
take block size $2$ so that on the renormalized site, $T^2$
is always equal to $I$. Using argument similar to the case
of on-site unitary symmetry, we can keep track and redefine
symmetry operations as we do renormalization. Finally, at
the fixed point we have a state described by matrices
$(A^{[k]}_{i^li^r})^{(\infty)}$ which is invariant under
time reversal operation
$(T^{[k]})^{(\infty)}=(u^{[k]})^{(\infty)}K$, that is,
\begin{align}
\label{Eqn:TR_A}
& \sum_{j^lj^r} u^{[k]}_{i^li^r,j^lj^r} (A^{[k]}_{j^lj^r})^*=
N_{[k]}^{-1}A^{[k]}_{i^li^r}M_{[k]}
\nonumber\\
& N_{[k]}= M_{[k-1]}
\end{align}
where the fixed-point label $\infty$ has been omitted. Solving this equation we find,\\
(a)$M_{[k]}M^*_{[k]}=e^{i\theta}I$. As $M_{[k]}$ is invertible, $e^{i\theta}=\pm 1$.\\
(b)$u^{[k]} = u^{[k],l} \otimes u^{[k],r}$. where $u^{[k],l}(u^{[k],l})^*=\pm I$ and $u^{[k],r}(u^{[k],r})^*=\pm I$.
Therefore, each entangled pair is time reversal invariant
\begin{equation}
(u^{[k],r}\otimes u^{[k+1],l})K |EP_{k,k+1}\rangle = |EP_{k,k+1}\rangle
\end{equation}
Similar to previous sections, we can show that $uu^*=I$ and $uu^*=-I$ correspond to two equivalence classes and two time reversal invariant fixed point states can be mapped into each other if and only if they belong to the same class. Therefore, our classification result for time reversal symmetry is
\[\frm{\emph{
For 1D gapped spin systems with ONLY time reversal symmetry,
there are two phases that do not break the symmetry.
}}
\]

If the system has additional translation symmetry, we can similarly classify the TI phases and find that
\[\frm{\emph{
For 1D systems with only translation and
time reversal symmetry $T$, there are two gapped phases
that do not break the two symmetries, if on each
site the time reversal transformation satisfies $T^2=I$.
}}
\]
1D integer-spin systems are examples of this case. The Haldane/AKLT state and the spin $0$ product state are representatives of the two phases respectively\cite{GW0931,PBT0959,PBT1039}.
We also have
\[\frm{\emph{
1D systems with translation and time reversal symmetry
are always gapless or have degenerate ground states, if on
each site the time reversal transformation satisfies $T^2=-I$.
}}
\]
1D half-integer-spin systems are examples of such case.

\section{Generalization to higher dimensions}
\label{highD}

In the last few sections, we classified symmetry protected
topological orders in one dimension, using (symmetric) LU
transformations. Can we use (symmetric) LU transformations
to classify (symmetry protected) topological orders in higher dimensions?

In higher dimensions, the situation is much more
complicated. First, infinitely many kinds of non-trivial
topological orders exist for class of systems without any
symmetries.\cite{Wrig,LWstrnet} A partial classification is
given in \Ref{CGW1035} for such a case in 2D. In the
presence of symmetry, the phase diagram is even more
complicated. \Ref{YK1065} studied 2D topological orders
protected by time reversal and point group symmetry.
So far, we do not have a detailed
understanding of topological phases in the presence of symmetry.

However, using similar arguments as that used for 1D systems,
we can obtain some simple partial results for higher dimensions.
For example, we have
\[\frm{\emph{
For $d$-dimensional spin systems with
only translation and an on-site symmetry $G$ which is
realized linearly, the object
$(\al,\om_1,\om_2,...,\om_d)$ label distinct gapped quantum
phases that do not break the two symmetries.  Here $\al$
labels the different 1D representations of
$G$ and $\om_i \in H^2(G,\C)$ label the
different types of projective representations of $G$.
}}
\]

\begin{figure}
\begin{center}
\includegraphics[scale=0.6]{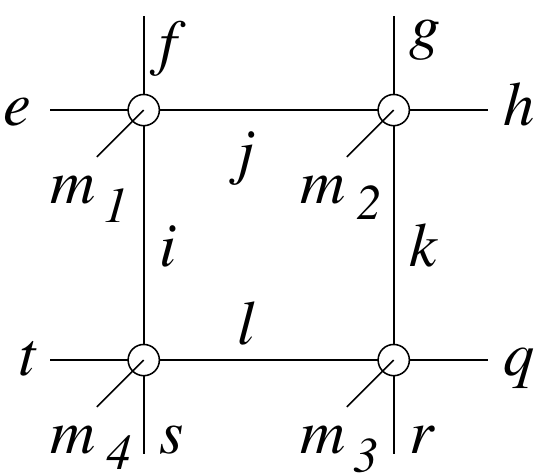}
\end{center}
\caption{Tensor-network -- a graphic representation of the
tensor-product wave function \eq{TNS} on
a 2D square lattice.  The indices on the links are summed over. }
\label{tps}
\end{figure}

\begin{figure}
\begin{center}
\includegraphics[scale=0.4]{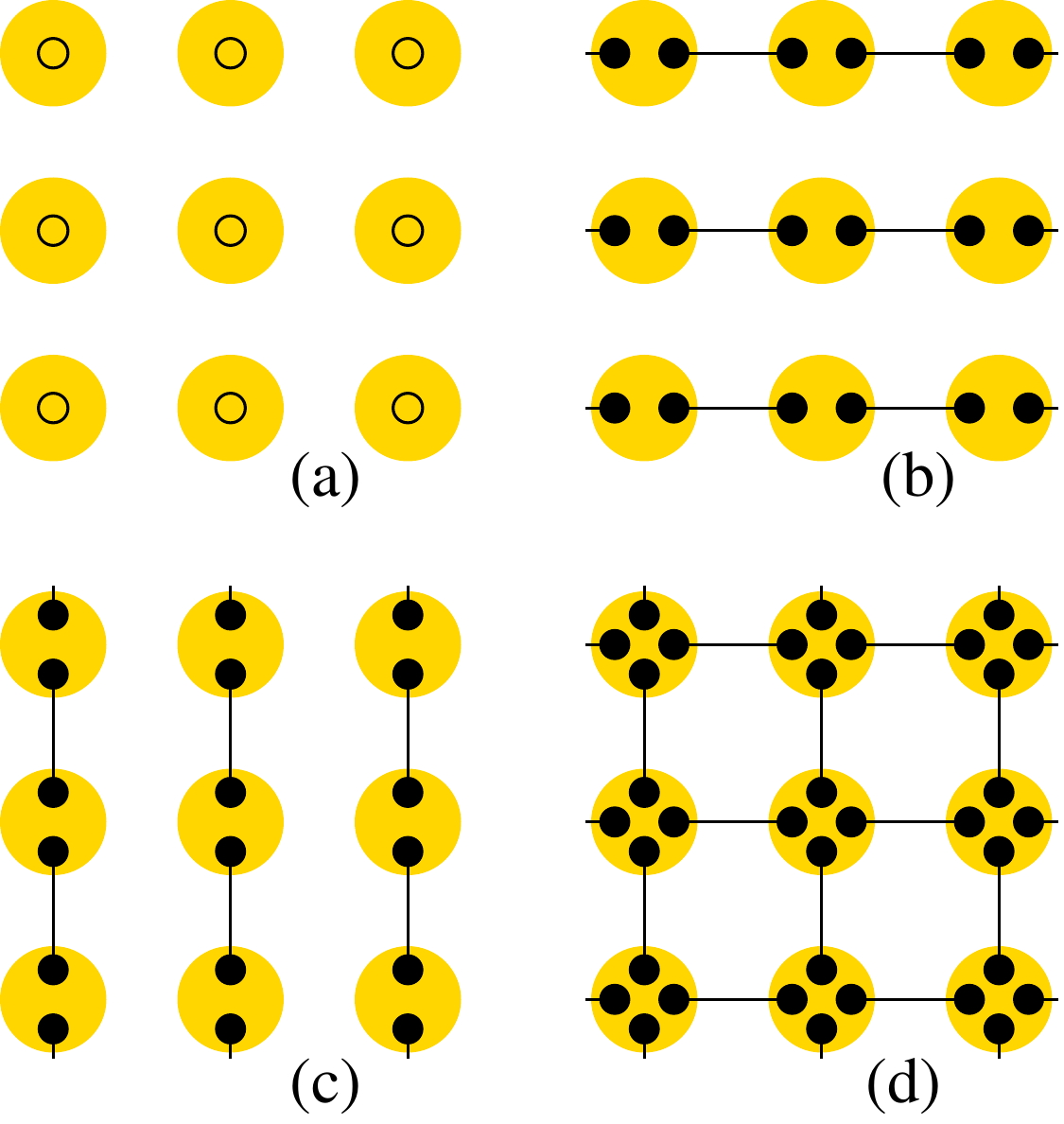}
\end{center}
\caption{
(Color online)
$( \om_1=0,1; \om_2=0,1)$ label four distinct states
in integer spin systems with translation and spin rotation
symmetries:
(a) $(\om_1,\om_2)=(0,0)$,
(b) $(\om_1,\om_2)=(0,1)$,
(c) $(\om_1,\om_2)=(1,0)$,
(d) $(\om_1,\om_2)=(1,1)$.
The open dots represent 0-spins.
The solid dots represent 1/2-spins.
Two solid dots connected by a line
represent a spin-singlet formed by two
1/2-spins.
The state in (a) has an spin-0 spin on each site.
The state in (b,c) has two spin-1/2 spins on each site
(or 4 states per site that form spin 0+1 representation
of $SO(3)$).
The state in (d) has four spin-1/2 spins on each site
(or 16 states per site that form spin 0+0+1+1+1+2 representation
of $SO(3)$).
}
\label{4states}
\end{figure}

 \begin{table*}[tb]
 \centering
 \begin{tabular}{ |c|c|c| }
 \hline
 Symmetry             & No. or Label of Different Phases & Example System\\ \hline
 None 		& 		1 		&  \\ \hline
 \multirow{2}{*}{On-site Linear Symmetry of Group G} &  \multirow{2}{*}{$\om \in H^2(G,\C)$ (*)}  & On-site $Z_n$ or $SU(2)$: 1 phase \\
                   &               & On-site $SO(3)$ or $D_2$ on integer spin: 2 phases \\ \hline
 \multirow{2}{*}{On-site Projective Symmetry of Group G}  & \multirow{2}{*}{$\om \in H^2(G,\C)$} & On-site $SO(3)$ or $D_2$ on \\
                &                                 & half-integer spin: 2 phases \\ \hline
 Time Reversal(TR)    &  2                              &          \\ \hline
 Translational Invariance(TI)    &  1                              &          \\ \hline
 TI+On-site           & \multirow{2}{*}{$\om \in H^2(G,\C)$ and $\alpha(G)$}  &  TI+On-site $Z_n$: n phases \\
 Linear Symmetry of Group G     &   &  TI+On-site $SO(3)$ on integer spin: 2 phases \\ \hline
 TI+ On-site          &   \multirow{2}{*}{0}  & TI+On-site $SO(3)$ or $D_2$ on\\
 Projective Symmetry of Group G &             & half-integer spin: no gapped phase \\ \hline
 TI+Parity             &   4                  &                              \\ \hline
 \multirow{2}{*}{TI+TR} & 2 if $T^2=I$  & TI+TR on integer spin: 2 phases     \\
                        & 0 if $T^2=-I$ & TI+TR on half-integer spin: no gapped phase \\ \hline
 \end{tabular}
 \caption{Summary of classification result for 1D gapped spin system with symmetric ground states. TI stands for translational invariance. TR stands for time reversal symmetry. $H^2(G,\C)$ is the second cohomology group of group $G$ over complex number $\C$. $\alpha(G)$ is a 1D representation of $G$. (*): this result applies when $\alpha(G)$ form a finite group, when $G=U(1)$, further classification according to different $\alpha(U(1))$ exist.}
 \label{Table:result}
 \end{table*}

Let us illustrate the above result in 2D by considering a
tensor network state (TNS) on a square lattice, where the
physical states living on each vertex $\v i$ are labeled by
$m_{\v i}$.  A translation invariant TNS is defined by the
following expression for the many-body wave function
$\Psi(\{m_{\v i}\})$:
\begin{eqnarray}
\Psi(\{ m_{\v i}\})=\sum_{ijkl\cdots}
A_{ejif}^{m_1}A_{jhkg}^{m_2}A_{lqrk}^{m_3}A_{tlsi}^{m_4}\cdots
\label{TNS}
\end{eqnarray}
Here $A_{ijkl}^{m_{\v i}}$ is a complex tensor
with one physical index $m_{\v i}$ and four
inner indices $i,j,k,l$. The physical index runs over
the number of physical states $d$ on each vertex, and the
inner indices runs over $D$ values. The TNS can be
represented graphically as in Fig. \ref{tps}.

If the tensor $A$ satisfy
\begin{align}
& A_{lrud}^{m}
= \al(g) \sum_{l'r'u'd'}
u_{mm'}(g) \times
\\
& \ \ \ \ \ \ (M^{-1}(g))_{ll'}
M_{r'r}(g)
(N^{-1}(g))_{dd'}
N_{u'u}(g)
A_{l'r'u'd'}^{m'},
\nonumber
\end{align}
for some invertible matrices $M(g)$ and $N(g)$,
then the many-body wave function $\Psi(\{ m_{\v i}\})$ is
symmetric under the on-site symmetry transformation $g$ in
the on-site symmetry group $G$.  Here $\al(g)$ is a one
dimensional representation of $G$, the $D$ by $D$ matrices
$M(g)$ form a projective representation represented by
$\om_1 \in H^2(G,\C)$, and the $D$ by $D$ matrices $N(g)$
form a projective representation represented by $\om_2 \in
H^2(G,\C)$.  Since symmetric LU transformation cannot change
$(\al,\om_1,\om_2)$, thus $(\al,\om_1,\om_2)$ label distinct
quantum phases.

In fact $(\al,\om_1,\om_2)$ are all measurable, so they
indeed label distinct quantum phases.  On a torus of size
$L_x\times L_y$, the symmetric many-body wave function
$\Psi(\{ m_{\v i}\})$ transforms as the 1D representation
$\al^{L_xL_y}(g)$ under the on-site symmetry transformation
$g$. If $G$ has only a finite number of 1D representations,
we can always choose $L_x$ and $L_y$ such that
$\al^{L_xL_y}(g)=\al(g)$.

On a cylinder of size $L_x\times L_y$ with open boundary in
the $x$-direction, the states on one boundary will form a
projective representation which is represented by $L_y \om_1
\in H^2(G,\C)$.  Similarly, if the open boundary in the
$y$-direction, the states on one boundary will form a
projective representation which is represented by $L_x \om_2
\in H^2(G,\C)$.  If we choose $L_x$ and $L_y$ properly (for
example to make $L_x \om_2=\om_2$ and $L_y \om_1=\om_1$), we
can detect both $\om_1$ and $\om_2$.

A system with integer on-site spins gives us an example with
$G=SO(3)$, if the system has translation and spin rotation
symmetry.  For $G=SO(3)$ there is no non-trivial 1D representations.
So we can drop the consideration of $\al$. Also $G=SO(3)$ has two types of
projective representations: $H^2(SO(3),\C)=\{0,1\}$, where
$\om=0$ is the trivial projective representation which
correspond to linear representations of $SO(3)$ (on integer
spins) and  $\om=1$ is the non-trivial projective
representation which corresponds to half-integer spins.  Thus
$(\om_1=0,1; \om_2=0,1)$ label four distinct states in 2D (see
Fig.  \ref{4states}).

\section{Conclusion}
\label{conc}

Using the (symmetric) Local Unitary equivalence relation
between gapped ground states in the same phase and the matrix product representation of 1D states, we classify possible quantum phases for strongly interacting 1D spin system with certain symmetry when ground state of the system does not break the symmetry. Our result is summarized in table \ref{Table:result}.

Many well known results are rederived using a quite
different approach, for example the existence of Haldane phase\cite{H8364} and the gaplessness of spin $1/2$ Heisenberg model\cite{LSM6107}. Those results are also greatly
generalized to new situations, for example to the case of
time reversal symmetry and $D_2$ symmetry. We find that the projective
representations play a very important role in understanding
and formulating those generalized results.

In higher dimensions, things are more complicated. Nevertheless, similar considerations allow us to obtain some interesting examples of symmetry protected topological orders. A complete classification of higher dimension phases, however, requires at least a full understanding of topological orders, an element that is absent in the 1D phase diagram.

Results similar to ours on the classification of integer/half-integer 1D spin
chains with SO(3) symmetry have been obtained independently by Alexei Kitaev
recently.\cite{Kitaev}

We would like to thank A. Kitaev, J. I. Cirac, F.  Verstraete, G. Vidal,
G.-M. Zhang, A. W.W. Ludwig and H. Katsura for some very helpful discussions.  XGW is
supported by  NSF Grant No.  DMR-1005541. ZCG is supported
in part by the NSF Grant No.  NSFPHY05-51164.

\appendix

\section{Local Unitary Transformation on Matrix Product State and Invariance of Double Tensor}
\label{dbt}

In this section we present the proof for the theorem that:
\textit{Two sets of matrices $A_{i,\alpha\beta}$ and
$B_{j,\alpha\beta}$, where $i$,$j$ label different matrices
and $\alpha\beta$ are the row and column indices of the
matrices, give rise to the same double tensor
$\mathbb{E}_{\alpha\gamma,\beta\chi}=\sum_{i}
A_{i,\alpha\beta} \times A^*_{i,\gamma\chi} = \sum_{j}
B_{j,\alpha\beta} \times B^*_{j,\gamma\chi}$, if and only
if they are related by a unitary transformation
$B_{i,\alpha\beta}=\sum_{j} U_{ij} A_{j,\alpha\beta}$.} The
dimension $M$ of $i$ and the dimension $N$ of $j$ can be
different. The $M\times N$ matrix $U$ is in general called
unitary if $U^{\dagger}U=UU^{\dagger}=I_L$, the identity
matrix in $L$ dimensional space where $L \le M$ and $L \le
N$. The complete proof of this theorem can be found in
\Ref{NieC00}, where this property is discussed in terms
of `the unitary degree of freedom in the operator sum
representation of quantum channels'. Here we re-present this
proof following the notation and terminology of the current
paper for simplicity of understanding.

First we prove the `if' part of the theorem. Suppose that
$B_{i,\alpha\beta}=\sum{j} U_{ij} A_{j,\alpha\beta}$, then
\begin{eqnarray}
\mathbb{E}^B_{\alpha\gamma,\beta\chi} &=&\sum_{i} B_{i,\alpha\beta} \times B^*_{i,\gamma\chi} \\ \nonumber
 &=& \sum_{i} \sum_{j_1} \sum_{j_2} U_{ij_1}A_{j_1,\alpha\beta} \times U^*_{ij_2}A^*_{j_2,\gamma\chi} \\ \nonumber
 &=& \sum_{j_1,j_2} \sum_{i} U_{ij_1}U^{\dagger}_{j_2i} A_{j_1,\alpha\beta} \times A^*_{j_2,\gamma\chi} \\ \nonumber
 &=& \sum_{j_1} A_{j_1,\alpha\beta} \times A^*_{j_1,\gamma\chi}
 = \mathbb{E}^A_{\alpha\gamma,\beta\chi}
\end{eqnarray}
Therefore the two double tensors are the same. This proves
the first part of the theorem.

On the other hand, suppose that the two double tensors are
the same $\mathbb{E}^A_{\alpha\gamma,\beta\chi} =
\mathbb{E}^B_{\alpha\gamma,\beta\chi} =
\mathbb{E}_{\alpha\gamma,\beta\chi}$. Reorder the indices
of $\mathbb{E}$ and treat it as a matrix with row indices
$\alpha\beta$ and column indices $\gamma\chi$. We will
denote the double tensor after this reordering as
$\hat{\mathbb{E}}_{\alpha\beta,\gamma\chi}$. It is easy
to see that $\hat{\mathbb{E}}$ is a positive semi-definite
matrix, as
\begin{align}
&
\sum_{\alpha\beta,\gamma\chi} v^*_{\alpha\beta} \hat{\mathbb{E}}_{\alpha\beta,\gamma\chi} v_{\gamma\chi}
\\ \nonumber
& =
\sum_{i} (\sum_{\alpha\beta} v^*_{\alpha\beta}B_{i,\alpha\beta}) \times (\sum_{\gamma\chi}B^*_{i,\gamma\chi}v_{\gamma\chi})
\geq 0
\end{align}
for any vector $v_{\alpha\beta}$

Diagonalize $\hat{\mathbb{E}}$ into
\begin{equation}
\hat{\mathbb{E}}_{\alpha\beta,\gamma\chi} = \sum_{k} \lambda_k e_{k,\alpha\beta} \times e^*_{k,\gamma\chi}
\end{equation}
with $\lambda_k \ge 0$. Define vectors
$\tilde{e}_{k,\alpha\beta}=\sqrt{\lambda_k}
e_{k,\alpha\beta}$, so that
$\hat{\mathbb{E}}_{\alpha\beta,\gamma\chi} = \sum_{k}
\tilde{e}_{k,\alpha\beta} \times
\tilde{e}^*_{k,\gamma\chi}$. $\tilde{e}_{k,\alpha\beta}$
form a complete orthogonal set and hence we can expand
$A_{i,\alpha\beta}$ and $B_{i,\alpha\beta}$ in terms of
them.
\begin{eqnarray}
A_{i,\alpha\beta}= \sum_k P_{ik} \tilde{e}_{k,\alpha\beta} \\ \nonumber
B_{i,\alpha\beta}= \sum_k Q_{ik} \tilde{e}_{k,\alpha\beta} \\ \nonumber
\end{eqnarray}
Then
\begin{eqnarray}
\hat{\mathbb{E}}_{\alpha\beta,\gamma\chi} &=& \sum_{i} A_{i,\alpha\beta} \times A^*_{i,\gamma\chi} \\ \nonumber
 &=& \sum_{k_1k_2} \sum_{i} P_{ik_1}P^*_{ik_2} \tilde{e}_{k_1,\alpha\beta} \times \tilde{e}^*_{k_2,\gamma\chi}
\end{eqnarray}
But we know that
$\hat{\mathbb{E}}_{\alpha\beta,\gamma\chi} = \sum_{k}
\tilde{e}_{k,\alpha\beta} \times
\tilde{e}^*_{k,\gamma\chi}$. Therefore, $\sum_{i}
P_{ik_1}P^*_{ik_2} = \delta_{k_1k_2}$ and $P$ is a unitary
matrix.

Similarly we can show that $Q$ is a unitary matrix.
Therefore $A_{i,\alpha\beta}$ and $B_{j,\alpha\beta}$ is
related by a unitary transformation $U_{ij}$ where
$U=PQ^{\dagger}$. We have thus proved both part of the
theorem.

\section{Degeneracy of Largest Eigenvalue of Double Tensor
and Correlation Length of Matrix Product State}
\label{ndg}

In section \ref{1DGapMPS}, we cited the property of
matrix product state that the finite correlation length of
the state is closely related to the non-degeneracy of the
largest eigenvalue of the double tensor, as discussed in
Ref. \onlinecite{FNW9243,PVW0701}. In this section, we give
a brief illustration of why this is so. For simplicity of
notation, we focus on the translational invariant case first.
Generalization to matrix product states without translational invariance is straightforward and similar conclusions can be reached.

For a matrix product state $| \phi \rangle$ described by
matrices $A_{i,\alpha\beta}$ with double tensor
$\mathbb{E}_{\alpha\gamma,\beta\chi}=\sum_{i}
A_{i,\alpha\beta} \times A^*_{i,\gamma\chi}$, define
$\mathbb{E}[O]_{\alpha\gamma,\beta\chi}=\sum_{ij} O_{ij}
A_{i,\alpha\beta} \times A^*_{j,\gamma\chi}$ for arbitrary
operator $O_{ij}$. Follow the previous convention and treat
$\mathbb{E}$ and $\mathbb{E}[O]$ as matrices with row index
$\alpha\gamma$ and column index $\beta\chi$. The norm of
the wavefunction is given by $\langle \phi | \phi \rangle =
\Tr (\mathbb{E}^N)$, where $N$ is total length of the chain.
WLOG, we can set the largest eigenvalue of $\mathbb{E}$ to
be $1$ and hence the norm goes to a finite value (dimension
of the eigenspace) as $N$ goes to infinity. The expectation
value of any local operator $O$ is $\langle O \rangle =
\Tr(\mathbb{E}^{N-1}\mathbb{E}[O])/\Tr (\mathbb{E}^N)$ and the correlation
between two operators $O_1$ and $O_2$ becomes
\begin{equation}
\begin{array}{l}
\langle O_1 O_2 \rangle -\langle O_1 \rangle \langle O_2 \rangle= \\
\Tr(\mathbb{E}^{N-L-2}\mathbb{E}[O_1]\mathbb{E}^{L}\mathbb{E}[O_2])/\Tr (\mathbb{E}^N)- \\
\Tr(\mathbb{E}^{N-1}\mathbb{E}[O_1])\Tr(\mathbb{E}^{N-1}\mathbb{E}[O_2])/\Tr^2 (\mathbb{E}^N)
\end{array}
\end{equation}

The physical constraints on the expectation value and
correlation functions of local operators require that the
double tensor $\mathbb{E}$ has certain properties. First,
put $\mathbb{E}$ into its Jordan normal form and decompose
it as $\mathbb{E}=\sum_{\lambda} \lambda
P_{\lambda}+R_{\lambda}$, where $P_{\lambda}$ is the
diagonal part and $R_{\lambda}$ the Nilpotent part. But for
the largest eigenvalue $1$, $R_1$ must be $0$ as otherwise
for large system size $N$ $\langle O \rangle =
\Tr(\mathbb{E}^{N-1}\mathbb{E}[O])/\Tr (\mathbb{E}^N) \sim \Tr((P_1+R_1)^N
\mathbb{E}[O])$ will be unbounded for any $\mathbb{E}[O]$
that satisfies $\Tr(R_1\mathbb{E}[O])\neq 0$. The physical requirement
that any local operator has bounded norm requires that $R_1$
must be $0$.

Next we will show that the dimension of $P_1$ is closely
related to the correlation length of the state. At large
system size $N$, the correlator $\langle O_1 O_2 \rangle
-\langle O_1 \rangle \langle O_2 \rangle =
\Tr(P_1\mathbb{E}[O_1](\sum_{\lambda}\lambda
P_{\lambda}+R_{\lambda})^{L} \mathbb{E}[O_2])/\Tr (P_1) -
\Tr(P_1\mathbb{E}[O_1])\Tr(P_1\mathbb{E}[O_2])/\Tr^2 (P_1)$. When $L$ is
large, we keep only the first order term in
$(\sum_{\lambda}\lambda P_{\lambda}+R_{\lambda})^{L}$ and
the correlator goes to
$\Tr(P_1\mathbb{E}[O_1]P_1\mathbb{E}[O_2])/\Tr (P_1)-\Tr(P_1\mathbb{E}[O_1])\Tr(P_1\mathbb{E}[O_2])/\Tr^2 (P_1)$.
If $P_1$ is one dimensional, the two terms both become
$\langle v_1|\mathbb{E}[O_1]| v_1\rangle\langle
v_1|\mathbb{E}[O_2]|v_1\rangle$ and cancel each other for
any $O_1,O_2$ and the second order term in
$(\sum_{\lambda}\lambda P_{\lambda}+R_{\lambda})^{L}$
dominates which decays as $\lambda^L$. For $\lambda<1$, the
correlator goes to zero exponentially and the matrix product
state as finite correlation length. On the other hand, if
$P_1$ is more than one dimensional, the first order term has
a finite contribution independent of $L$ $\sum_{i,j}
\langle v_i|\mathbb{E}[O_1]| v_j\rangle\langle
v_j|\mathbb{E}[O_2]|v_i\rangle/\Tr(P_1) - \langle
v_i|\mathbb{E}[O_1]|v_i \rangle \langle
v_j|\mathbb{E}[O_2]|v_j\rangle/\Tr^2(P_1)$, where $v_i,v_j$ are
eigenbasis for $P_1$. Therefore, degeneracy of the largest
eigenvalue of the double tensor implies non-decaying
correlation. To describe quantum states with finite
correlation length, the double tensor must have a largest
eigenvalue which is non-degenerate.

If the system is not translational invariant and $\mathbb{E}^{[k]}$ vary from site to site, we cannot diagonalize all the double tensors at the same time. However, as shown in the `canonical form' of \Ref{PVW0701}, there is a largest eigenspace of $\mathbb{E}^{[k]}$(with eigenvalue $1$) such that the right eigenvector on site $k$ is the same as the left eigenvector on site $k+1$. Therefore, when multiplied together this eigenspace will always have eigenvalue $1$. There could be other eigenspaces with eigenvalue $1$ and matching eigenvectors from site to site. However, then we can show similar to the TI case that this leads to an infinite correlation length. On the other hand, other eigenspaces could have eigenvalues smaller than $1$ or they have mis-matched eigenvectors. If this is the case, all other eigenspaces decay exponentially with the number of sites multiplied together which gives rise to a finite correlation length. We will say that $\mathbb{E}^{[k]}$ has a non-degenerate largest eigenvalue $1$ for this case in general.

\section{Projective Representation}
\label{prorep}

Operators $u(g)$ form a projective representation of symmetry group $G$ if
\begin{align}
 u(g_1)u(g_2)=\om(g_1,g_2)u(g_1g_2),\ \ \ \ \
g_1,g_2\in G.
\end{align}
Here $\om(g_1,g_2) in \C$, the factor system of the projective representation, satisfies
\begin{align}
 \om(g_2,g_3)\om(g_1,g_2g_3)&=
 \om(g_1,g_2)\om(g_1g_2,g_3),
\end{align}
for all $g_1,g_2,g_3\in G$.
If $\om(g_1,g_2)=1$, this reduces to the usual linear representation of $G$.

A different choice of pre-factor for the representation matrices
$u'(g)= \bt(g) u(g)$ will lead to a different factor system
$\om'(g_1,g_2)$:
\begin{align}
\label{omom}
 \om'(g_1,g_2) =
\frac{\bt(g_1g_2)}{\bt(g_1)\bt(g_2)}
 \om(g_1,g_2).
\end{align}
We regard $u'(g)$ and $u(g)$ that differ only by a pre-factor as
equivalent projective representations and the corresponding
factor systems $\om'(g_1,g_2)$ and $\om(g_1,g_2)$ as
belonging to the same class $\om$.

Suppose that we have one projective representation $u_1(g)$
with factor system $\om_1(g_1,g_2)$ of class $\om_1$ and
another $u_2(g)$ with factor system $\om_2(g_1,g_2)$ of
class $\om_2$, obviously $u_1(g)\otimes u_2(g)$ is a
projective presentation with factor group
$\om_1(g_1,g_2)\om_2(g_1,g_2)$. The corresponding class
$\om$ can be written as a sum $\om_1+\om_2$. Under such an
addition rule, the equivalence classes of factor systems
form an Abelian group, which is called the second cohomology
group of $G$ and denoted as $H^2(G,\C)$. The identity
element $\om_0$ of the group is the class that contains the linear representation of the group.

Here are some simple examples:\\
(a) cyclic groups $Z_n$ do not have non-trivial projective representation. Hence for $G=Z_n$, $H^2(G,\C)$ contains only the identity element.\\
(b) a simple group with non-trivial projective representation is the Abelian dihedral group $D_2=Z_2\times Z_2$. For the four elements of the group $(0/1,0/1)$, consider representation with Pauli matrices $g(0,0)=\begin{bmatrix} 1 & 0 \\ 0 & 1 \end{bmatrix}$, $g(0,1) = \begin{bmatrix} 0 & 1 \\ 1 & 0 \end{bmatrix}$, $g(1,0) = \begin{bmatrix} 1 & 0 \\ 0 & -1 \end{bmatrix}$, $g(1,1) = \begin{bmatrix} 0 & -i \\ i & 0 \end{bmatrix}$. It can be check that this gives a non-trivial projective representation of $D_2$. \\
(c) when $G=SO(3)$, $H^2(G,\C)=Z_2$. The two elements
correspond to integer and half-integer representations of
$SO(3)$ respectively.\\
(d) when $G=U(1)$, $H^2(G,\C)$ is trivial:
$H^2(U(1),\C)=Z_1$.  We note that $\{ \e^{\imth m \th} \}$
form a representation of $U(1)=\{ \e^{\imth \th} \}$ when
$m$ is an integer.  But $\{ \e^{\imth m \th} \}$ will form
a projective representation of $U(1)$ when $m$ is not an
integer.  But under the equivalence relation \eq{omom}, $\{
\e^{\imth m \th} \}$ correspond to the trivial projective
representation, if we choose $\bt(g)=\e^{-\imth m \th}$. Note
that $\bt(g)$ can be a discontinuous function over the group
manifold.

\section{Solving symmetry condition for fixed point}
\label{soluu}

In this section, we explicitly solve the symmetry condition
\eqn{Eqn:LU_A}. The goal is to 1. classify possible
symmetry operations at fixed point and 2. find the corresponding symmetric
fixed point state.
For simplicity, we
drop the site index $[k]$ and rewrite \eqn{Eqn:LU_A} as
\begin{equation}
\label{Eqn:LU_A_C9}
\sum_{j^lj^r} \frac{u_{i^li^r,j^lj^r}(g)}{\alpha^{(R)}(g)} A_{j^lj^r}=
N^{-1}(g)A_{i^li^r}M(g)
\end{equation}
where $u(g)$ is a projective or linear unitary representation of $G$,
the matrix $A_{i^li^r}$ is given by its matrix elements
$A_{i^li^r,\alpha\beta}=\sqrt{\lambda^l_{i^l}}\delta_{i^l\alpha}
\cdot \sqrt{\lambda^r_{i^r}}\delta_{i^r\beta}$ with
$i^l,\alpha=1,...,D_l$,
$i^r,\beta=1,...,D_r$,
and $M(g),\ N(g)$ are sets of
invertible matrices labeled by $g$.
Since $\frac{u(g)}{\alpha^{(R)}(g)}$
is also a projective or linear unitary representation of
$G$, we can
absorb $\alpha^{(R)}(g)$ into $u(g)$ and rewrite
\eqn{Eqn:LU_A_C9} as
\begin{equation}
\label{Eqn:LU_A_C}
\sum_{j^lj^r} u_{i^li^r,j^lj^r}(g) A_{j^lj^r}=
N^{-1}(g)A_{i^li^r}M(g)
\end{equation}
We note that matrix elements $A_{i^li^r,\al\bt}$ is
non-zero only when $\alpha=i^l$, $\beta=i^r$ and the
full set of $\{A_{i^li^r}\}$ form a complete basis in the
space of $D_l\times D_r$ dimensional matrices.

$M(g),\ N(g)$ do not necessarily form a representation of $G$.
But the fixed point form of the matrices requires that $M(g),\
N(g)$ be a projective representation, as on the one hand
\begin{align}
 & \sum_{j^lj^r} u_{i^li^r,j^lj^r}(g_1g_2) A_{j^lj^r} \\ \nonumber
=& \sum_{j^lj^rk^lk^r}
\om_{sym}(g_1,g_2)
u_{i^li^r,k^lk^r}(g_1)u_{k^lk^r,j^lj^r}(g_2) A_{j^lj^r} \\ \nonumber
=& \sum_{k^lk^r}
\om_{sym}(g_1,g_2)
u_{i^li^r,k^lk^r}(g_1) N^{-1}(g_2)A_{k^lk^r}M(g_2) \\ \nonumber
=&
\om_{sym}(g_1,g_2)
N^{-1}(g_2)N^{-1}(g_1)A_{i^li^r}M(g_1)M(g_2) \\ \nonumber
\end{align}
and on the other hand
\begin{align}
 & \sum_{j^lj^r} u_{i^li^r,j^lj^r}(g_1g_2) A_{j^lj^r} \\ \nonumber
=& N^{-1}(g_1g_2)A_{i^li^r}M(g_1g_2)
\end{align}
Therefore
\begin{align}
&\ \ \
\om_{sym}(g_1,g_2)
N^{-1}(g_2)N^{-1}(g_1)A_{i^li^r}M(g_1)M(g_2)
\nonumber\\
&=N^{-1}(g_1g_2)A_{i^li^r}M(g_1g_2)
\end{align}
for all $i^li^r$. However, the set of matrices
$\{A_{i^li^r}\}$ form a complete basis in the space of
$D_l\times D_r$ dimensional matrices. Therefore,
\begin{align}
&\ \ \
\om_{sym}(g_1,g_2)
 N^{-1}(g_2)N^{-1}(g_1)\otimes M(g_1)M(g_2)
\nonumber\\
& =N^{-1}(g_1g_2)\otimes M(g_1g_2),
\end{align}
and
$N(g)$ and $M(g)$ form two projective representations
\begin{align}
\label{MNproj}
 N(gh)&=\om_N(g,h) N(g) N(h),
\nonumber\\
 M(gh)&=\om_M(g,h) M(g) M(h),
\end{align}
with $ |\om_N(g_1,g_2)|= |\om_M(g_1,g_2)|=1$ and
\begin{align}
\label{symMN}
 \om_{sym}(g_1,g_2)=
 \frac{\om_M(g_1,g_2)}{ \om_N(g_1,g_2)}
\end{align}

Let us rewrite \eqn{Eqn:LU_A_C} as
\begin{equation}
\label{Eqn:LU_A_C1}
N(g)(\sum_{j^lj^r} u_{i^li^r,j^lj^r}(g) A_{j^lj^r}) M^{-1}(g)=
A_{i^li^r}
\end{equation}
We note that
\begin{equation}
\label{Eqn:LU_A_C2}
N(g)(\sum_{j^lj^r}
(\t N^{-1})_{j^l,i^l}
\t M_{i^r,j^r}
A_{j^lj^r}) M^{-1}(g)=
A_{i^li^r}
\end{equation}
where the matrices $\t M$ and $\t N$ are given by
\begin{align}
\t M_{\al\bt} =   M_{\al\bt}
\frac
{\sqrt{\la^r_\al}}
{\sqrt{\la^r_\bt}}
,\ \ \ \
\t N_{\al\bt} =   N_{\al\bt}
\frac
{\sqrt{\la^l_\bt}}
{\sqrt{\la^l_\al}}
 .
\end{align}
Since the set of matrices $\{A_{i^li^r}\}$ form a complete
basis in the space of $D_l\times D_r$ dimensional matrices,
we find
\begin{align}
\label{uMN}
u_{i^li^r,j^lj^r}(g)=
(\t N^{-1})_{j^l,i^l}(g)
\t M_{i^r,j^r}(g) .
\end{align}
Putting back the factor of $\al^{(R)}(g)$, we find that
\begin{align}
\label{uaMN}
u_{i^li^r,j^lj^r}(g)= \al^{(R)}(g)
(\t N^{-1})_{j^l,i^l}(g)
\t M_{i^r,j^r}(g) .
\end{align}

\section{Equivalence Between Symmetric Fixed Point States}
\label{equiv}
From the solution in section \ref{soluu}, we know that the fixed point state symmetric under linear on-site symmetry of group $G$ takes the form
\begin{equation}
|\phi\rangle^{(\infty)} = |EP_{1,2}\rangle |EP_{2,3}\rangle ... |EP_{k,k+1}\rangle ...
\end{equation}
where $|EP_{k,k+1}\rangle$ is an entangled pair between the right virtual qubit on site $k$ and the left virtual qubit on site $k+1$(see Fig. \ref{fig:FP} upper layer). Each entangled pair is invariant under a linear symmetry transformation of the form $u^{[k],r}(g)\otimes u^{[k+1],l}(g)$
\begin{equation}
u^{[k],r}(g)\otimes u^{[k+1],l}(g) |EP_{k,k+1}\rangle = |EP_{k,k+1}\rangle
\end{equation}
But $u^{[k],r}(g)$ or $u^{[k+1],l}(g)$ alone might not form
a linear representation of $G$. They could in general be a
projective representation of $G$. If $u^{[k],r}(g)$ is a
projective representation corresponding to class $\om$ in
$H^2(G,\C)$, then $u^{[k+1],l}$ must correspond to class
$-\om$. $\om$ does not vary from site to site and labels a particular symmetric fixed point state.

Now we will show that symmetric fixed point states with the
same $\om$ can be connected through symmetric LU
transformations and hence belong to the same phase while
those with different $\om$ cannot and belong to different phases.

First, suppose that two symmetric fixed point states
$|\phi_1\rangle$ and $|\phi_2\rangle$ are related with the
same $\om$, i.e.
\begin{align}
u^{[k],r}_1(g)\otimes u^{[k+1],l}_1(g) |EP_{k,k+1}\rangle_1 = |EP_{k,k+1}\rangle_1 \\ \nonumber
u^{[k],r}_2(g)\otimes u^{[k+1],l}_2(g) |EP_{k,k+1}\rangle_2 = |EP_{k,k+1}\rangle_2
\end{align}
where $|EP_{k,k+1}\rangle_{1(2)}$ is an entangled pair of
virtual spins on Hilbert space $\mathcal{H}^{[k],r}_{1(2)}
\otimes \mathcal{H}^{[k+1],l}_{1(2)}$. $u^{[k],r}_{1(2)}(g)$
is a projective representation of $G$ corresponding to $\om$
on $\mathcal{H}^{[k],r}_{1(2)}$ and $u^{[k+1],l}_{1(2)}(g)$
a projective representation corresponding to $-\om$ on $\mathcal{H}^{[k+1],l}_{1(2)}$.

We can think of $|EP_{k,k+1}\rangle_1$ and $|EP_{k,k+1}\rangle_2$ as living together in a joint Hilbert space $(\mathcal{H}^{[k],r}_{1} \oplus \mathcal{H}^{[k],r}_{2})\otimes(\mathcal{H}^{[k+1],l}_{1} \oplus \mathcal{H}^{[k+1],l}_{2})$.  The symmetry representation on this joint Hilbert space can be defined as
\begin{align}
 & u^{[k],r}(g) \otimes u^{[k+1],l}(g)  \\ \nonumber
= & (u^{[k],r}_1(g) \oplus u^{[k],r}_2(g))\otimes(u^{[k+1],l}_1(g) \oplus u^{[k+1],l}_2(g))
\end{align}
As $u^{[k],r}_1(g)$ and $u^{[k],r}_2(g)$ (also
$u^{[k+1],l}_1(g)$ and $u^{[k+1],l}_2(g)$) both correspond
to $\om$ ($-\om$), their direct sum
$u^{[k],r}(g)$($u^{[k+1],l}(g)$) is also a projective
representation corresponding to $\om$($-\om$). Therefore, we have a linear representation of $G$ on each site $k$, $u^{[k],l}(g) \otimes u^{[k],r}(g)$ and both $|EP_{k,k+1}\rangle_1$ and $|EP_{k,k+1}\rangle_2$ are symmetric under $u^{[k],r}(g) \otimes u^{[k+1],l}(g)$.

\begin{figure}[t]
\begin{center}
\includegraphics[width=3.0in]{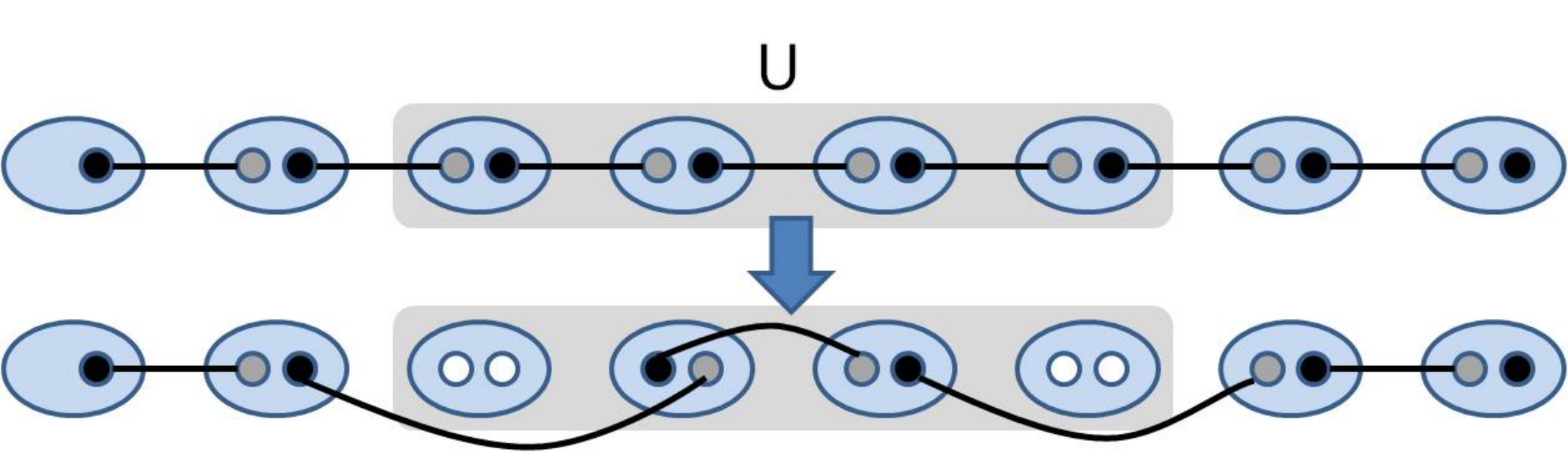}
\end{center}
\caption{
(Color online) Fixed point state related to projective
representation of class $\om$ before (upper) and after (lower)
a local unitary operation on the shaded region that does not
break the symmetry. White dots correspond to $\om_0$, the
identity element in $H^2(G,\C)$, black dots correspond to
$\om$ and gray dots correspond to $-\om$.}
\label{fig:inequiv}
\end{figure}

Now we can perform a LU transformation on the joint Hilbert space and rotate continuously between $|EP_{k,k+1}\rangle_1$ and $|EP_{k,k+1}\rangle_2$. That is,
\begin{equation}
U(\theta)= \cos(\frac{\theta}{2})I-i\sin(\frac{\theta}{2})(|a\rangle\langle b| +|b\rangle\langle a| )
\end{equation}
where $|a\rangle = |EP_{k,k+1}\rangle_1$, $|b\rangle =
|EP_{k,k+1}\rangle_2$ and $\theta$ goes from $0$ to $\pi$.
By doing this locally to each pair, we can map
$|\phi_1\rangle$ to $|\phi_2\rangle$ (and vice verse) with
LU transformations without breaking the on-site symmetry of
group $G$. Therefore, $|\phi_1\rangle$ and $|\phi_2\rangle$
belong to the same phase if they are related with the same
$\om$.

On the other hand, if $|\phi_1\rangle$ and $|\phi_2\rangle$
are related to $\om_1$ and $\om_2$ respectively, we will show that they cannot be connected by any LU transformation that does not break the symmetry.

Suppose that $\om_1$ is non-trivial, we start with $|\phi_1\rangle$ and apply a local unitary operation $U$ to a finite region (shaded in Fig. \ref{fig:inequiv}). $|\phi_1\rangle$ is composed of invariant singlets of symmetry group $G$. If $U$ does not break the symmetry, the resulting state should still be composed of singlets. The singlet pairs outside of the shaded region are not changed while those overlapping with the shaded region can take any possible structure after the operation $U$.

No matter what the change is, the right virtual spin on the
site to the left of the region corresponding to $\om_1$
should form a singlet with some degrees of freedom in the
region. As the singlet is invariant under a linear
representation of $G$, these degrees of freedom in the
region must form a projective representation of $G$
corresponding to $-\om_1$. These degrees of freedom could
live on one site or distribute over several sites. However,
the sites only support linear representations. Therefore,
there must be some remaining degrees of freedom on the same
sites that correspond to $\om_1$. These remaining degrees of
freedom must form singlets again with other degrees of
freedoms in the region that correspond to $-\om_1$. We
can continue this argument until finally some degree of
freedom in the region corresponding to $-\om_1$ connect with
the left virtual spin on the site to the right of the region
corresponding to $\om_1$ and form a singlet.

In Fig. \ref{fig:inequiv}, we illustrate one possible
structure of singlets after operation $U$. White dots
correspond to $\om_0$, the identity element in $H^2(G,\C)$,
black dots correspond to $\om_1$ and gray dots correspond to
$-\om_1$.

Therefore, we can see that no matter what the symmetric LU
operation might be on $|\phi_1\rangle$, singlet entangled
pairs related to $\om_1$ must connect head to tail and cover
the whole length of the chain. In other words, we can not
shrink a chain of singlet entangled pairs related to
non-trivial $\om_1$ continuously to a point or change it to
$\om_2$ by acting on it locally and without breaking the
symmetry. Hence fixed point states with different $\om$ cannot be related to each other with symmetric LU transformation and hence belong to different classes.

\section{A proof of \eqn{u0act_TI}}
\label{alL}

A gapped TI state can be represented by a uniform MPS.
After $R$ steps of $\{n_i\}$-block RG transformation,
we
obtain a MPS described by matrices
$(A_{i^li^r})^{(R)}$ which is
given by \eqn{Avv}.
To describe a state that does not break the on-site linear
symmetry, here $(A_{i^li^r})^{(R)}$ is
invariant (up to a phase) under
$u^{(R)}(g)$ on each site.
Therefore\cite{GWS0802},
\begin{align}
\label{Eqn:LU_ATI}
 \sum_{j^lj^r} u_{i^li^r,j^lj^r}(g) A_{j^lj^r}
& =
\al^{(R)}(g)
M^{-1}(g)A_{i^li^r} M(g)
\end{align}
must be satisfied with some invertible matrix $M(g)$.
Here we have dropped the RG step label $R$ (except in
$\al^{(R)}(g)$).  Each coarse grained site is a
combination of $\prod_{i=1}^R n_i$ original lattice sites
and $\al^{(R)}(g)$ form a 1D representation of $G$.

So if the number of sites
has a form $L=Q \prod_{i=1}^R n_i$, then
$\al_L(g)$ in \eqn{u0act} will have a form
\begin{align}
\label{alLIJ}
\al_L(g) = [\al^{(R)}(g) ]^Q
\end{align}
for any value of $Q$.
Now let us choose $Q=\prod_{i=1}^{R'} n'_i$
where $\prod_{i=1}^{R} n_i$ and $\prod_{i=1}^{R'} n'_i$
have no common factors. The total system size becomes
$L=\prod_{i=1}^R n_i \prod_{i=1}^{R'} n'_i$.
We can perform, instead, a
$R'$ step of $\{n'_i\}$-block RG transformation, which leads
to a 1D representation  $\al^{(R')}(g)$.
We find that
$\al_L(g)$ in \eqn{u0act} will have a form
\begin{align}
\al_L(g) = [\al^{(R')}(g) ]^{Q'}
\end{align}
where $Q'=L/\prod_{i=1}^{R'} n'_i=\prod_{i=1}^R n_i$.
Thus
\begin{align}
\al_L(g) =[\al^{(R)}(g) ]^{ \prod_{i=1}^{R'} n'_i}=
[\al^{(R')}(g) ]^{ \prod_{i=1}^{R} n_i} .
\end{align}
Since $\prod_{i=1}^{R} n_i$ and $\prod_{i=1}^{R'} n'_i$
have no common factors,
there must exist a 1D representation $\al(g)$ of $G$,
such that
\begin{align}
 \al^{(R)}(g) = [\al(g)]^{ \prod_{i=1}^{R} n_i}
, \ \ \
 \al^{(R')}(g) = [\al(g)]^{ \prod_{i=1}^{R'} n'_i}.
\end{align}
Now \eqn{alLIJ} becomes
\begin{align}
 \al_L(g) =  [\al(g)]^{Q \prod_{i=1}^{R} n_i}
=[\al(g)]^L
\end{align}
which gives us \eqn{u0act_TI}.

\section{Equivalence between gapped TI b-phases
and gapped TI phases}
\label{fullTI}

We have been using the quantum circuit formulation to study
TI systems and classify b-phases. As the quantum circuit
explicitly breaks translational symmetry, it is possible
that each b-phase contains several different TI phases. (On
the other hand, states in different b-phases must belong to
different TI phases.) In this section, we will show that
each b-phase actually corresponds to a single TI phase by
establishing a TI LU transformation between states in the
same b-phase. We will use the time evolution formulation of
LU transformation \eq{LU_H} and find a smooth path of
gapped TI Hamiltonian whose adiabatic evolution connects two
states within the same b-phase.

First, as an example, we consider the case of TI only and show that there is only one gapped TI phase. Each TI SRC MPS is described(up to local change of basis) by a double tensor $\mathbb{E}$ which has a non-degenerate largest eigenvalue $1$. $\mathbb{E}$ can be written as
\begin{equation}
\mathbb{E}_{\alpha\gamma,\beta\chi} = \mathbb{E}^0_{\alpha\gamma,\beta\chi} + \mathbb{E}'_{\alpha\gamma,\beta\chi} = \Lambda_{\alpha\gamma}\Lambda_{\beta\chi}+\mathbb{E}'_{\alpha\gamma,\beta\chi}
\label{E=E0+E'}
\end{equation}
where $\Lambda$ is the eigenvector of eigenvalue $1$ and $\mathbb{E}'$ is of eigenvalue less than $1$. In the `canonical form'\cite{PVW0701}, $\Lambda_{\alpha\gamma}=\lambda_{\alpha}\delta_{\alpha\gamma}$, $\lambda_{\alpha}>0$. Obviously, $\mathbb{E}^0$ is a valid double tensor and represents a state in fixed point form.
We can smoothly change $\mathbb{E}$ to $\mathbb{E}^0$ by turning down the $\mathbb{E}'$ term to $0$ from $t=0$ to $t=T$ as
\begin{equation}
\mathbb{E}(t)=\mathbb{E}^0+(1-\frac{t}{T})\mathbb{E}'
\label{Et}
\end{equation}
Every $\mathbb{E}(t)$ represents a TI SRC MPS state. To see this, note that if we recombine the indices $\alpha\beta$ as row index and $\gamma\chi$ as column index and denote the new matrix as $\hat{\mathbb{E}}$, then both $\hat{\mathbb{E}}$ and $\hat{\mathbb{E}}^0$ are positive semidefinite matrices. But then every $\hat{\mathbb{E}}(t)$ is also positive semidefinite, as for any vector $|v\rangle$
\begin{equation}
\begin{array}{lll}
\langle v|\hat{\mathbb{E}}(t)|v\rangle & = & \langle v|\hat{\mathbb{E}}^0|v\rangle +(1-\frac{t}{T})\langle v|\hat{\mathbb{E}'}|v\rangle \\ \nonumber
 & = & (1-\frac{t}{T})\langle v|\hat{\mathbb{E}}|v\rangle + \frac{t}{T} \langle v|\hat{\mathbb{E}}^0|v\rangle > 0
\end{array}
\end{equation}
$\mathbb{E}(t)$ is hence a valid double tensor and the state represented can be determined by decomposing $\mathbb{E}(t)$ back into matrices $A_i(t)$. Such a decomposition is not unique but WLOG, we can fix the decomposition scheme, so that $A_i(t)$ vary continuously with time and reach the fixed point form at $t=T$(up to local change of basis). The state represented $|\phi(t)\rangle$ hence also changes smoothly with $t$ and has a finite correlation length as all eigenvalues of $\mathbb{E}(t)$ expect for $1$ are diminishing with $t$. Therefore, $\mathbb{E}(t)$ represents a smooth path in TI SRC MPS that connects any state to a fixed point state(up to local change of basis).

How do we know that no phase transition happens along the
path? This is because for every state $|\phi(t)\rangle$, we
can find a parent Hamiltonian which changes smoothly with
$t$ and has the state as a unique gapped ground
state.\cite{SC10} Following the construction in
\Ref{FNW9243,PVW0701}, we choose a sufficiently large but
finite $l$ and set the parent Hamiltonian to be
$H(t)=-\sum_k h(t)_{k,k+l}$, where $h(t)_{k,k+l}$ is the
projection onto the support space of the reduced density
matrix on site $k$ to $k+l$ at time $t$. Note that this
Hamiltonian is translation invariant. For large enough $l$,
$h(t)_{k,k+l}$ will always be $D\times D$ dimensional. As
the state changes continuously, its reduced density matrices
of site $k$ to $k+l$ changes smoothly. Because the dimension
of the space does not change, $h(t)_{k,k+l}$ also changes
smoothly with time. Moreover, it can be shown that $H(t)$ is
always gapped as the second largest eigenvalue of
$\mathbb{E}(t)$ never approaches $1$ \cite{FNW9243,PVW0701}.
Therefore, by evolving the Hamiltonian adiabatically from
$t=0$ to $t=T$, we obtain a local unitary
transformation\cite{HW0541} connecting any state to the
fixed point form, and in particular without breaking the
translation symmetry.

Because any TI fixed point state can be disentangled into
product state in a TI way, we find that \emph{all TI 1D
gapped ground states are in the same phase, if no other
symmetries are required.}

If the system is TI and has on-site symmetry, we need to maintain the on-site symmetry while doing the smooth deformation. A TI SRC MPS which is symmetric under on-site symmetry of group $G$ is described by matrices which satisfy
\begin{equation}
\sum_j u_{ij}(g)A_j = \alpha(g) M^{-1}(g)A_iM(g)
\end{equation}
for some invertible projective representation $M(g)$.
The double tensor $\mathbb{E}$ hence satisfies
\begin{align}
\mathbb{E}_{\alpha\gamma,\beta\chi}=
\sum_{\alpha'\beta'\gamma'\chi'} M^{-1}_{\alpha\alpha'} M_{\beta\beta'} (M^*)^{-1}_{\gamma\gamma'} M^*_{\chi\chi'} \mathbb{E}_{\alpha'\gamma',\beta'\chi'}
\label{symm_E}
\end{align}
where the group element label $g$ has been omitted.
Being the non-degenerate one-dimensional eigenspace of $\mathbb{E}$, $\mathbb{E}^0$ must be invariant under the same transformation, and so does $\mathbb{E}'$. Therefore we have
\begin{align*}
\mathbb{E}^0_{\alpha\gamma,\beta\chi} = \sum_{\alpha'\beta'\gamma'\chi'} M^{-1}_{\alpha\alpha'} M_{\beta\beta'} (M^*)^{-1}_{\gamma\gamma'} M^*_{\chi\chi'} \mathbb{E}^0_{\alpha'\gamma',\beta'\chi'}
\\ \nonumber
\mathbb{E}'_{\alpha\gamma,\beta\chi} = \sum_{\alpha'\beta'\gamma'\chi'} M^{-1}_{\alpha\alpha'} M_{\beta\beta'} (M^*)^{-1}_{\gamma\gamma'} M^*_{\chi\chi'} \mathbb{E}'_{\alpha'\gamma',\beta'\chi'}
 \nonumber
\end{align*}
Now we smoothly change the double tensor as in Eqn.\ref{Et}. Evidently, the symmetry condition Eqn.\ref{symm_E} is satisfied for all $t$.

Decompose $\mathbb{E}(t)$ back to matrices $A_i(t)$ so that the represented state $|\phi(t)\rangle$ changes smoothly with time. Denote the symmetry transformed double tensor as $\mathbb{E}_{M(g)}$. As $\mathbb{E}_{M(g)}(t) = \mathbb{E}(t)$, there must exist a unitary operator $\t u(g)(t)$, such that
\begin{equation}
\sum_j \t u_{ij}(g)(t)A_j(t)= M^{-1}(g)A_i(t)M(g)
\end{equation}
where $\t u(g)(t)$ is a linear representation of $G$. Redefine $u(g)(t)=\t u(g)(t)\times \alpha(g)$, then
\begin{equation}
\sum_j u_{ij}(g)(t)A_j(t)= \alpha(g) M^{-1}(g)A_i(t)M(g)
\end{equation}
As $A_i(t)$ is chosen to be continuous with time, from the above equation we can see that $u(g)(t)$ is also continuous in time. On the other hand, $u(g)(t)$ form a linear representation of $G$. For all the cases we are interested in, the linear representations of $G$ are discrete. Therefore, as $u(g)(t)$ evolves smoothly with time, it cannot change from one representation to another but only from one equivalent form to another which differ by a unitary conjugation. That is, $u(g)(t)=V(t)u(g)V^{\dagger}(t)$, with a continuous $V(t)$. We can incorporate $V(t)$ into the matrices $A_i(t)$ and define $\t A_i(t) = \sum_j V^{\dagger}_{ij}(t) A_j(t)$, so that $\t A_i(t)$ is symmetric under $u(g)$ for all $t$. In the following discussion, we will assume that such a redefinition is made and the symmetry operation of the system will always be $u(g)\otimes...\otimes u(g)$.
Therefore, the continuous evolution of $\mathbb{E}(t)$ from $t=0$ to $t=T$ corresponds to a continuous evolution of short range correlated states $|\phi(t)\rangle$ which is always symmetric under the same on-site symmetry $u(g)$, with the same phase factor $(\alpha(g))^L$
and related to the same projective representation $\om$.

Such a smooth path in symmetric state space corresponds to a smooth path in symmetric Hamiltonian space. Construct parent Hamiltonian as discussed previously. Because the state is symmetric under on-site $u(g)$, the support space on site $k$ to $k+l$ must then form a representation space for $(\otimes u(g))^l$. Therefore, it is easy to see that the parent Hamiltonian, being a summation of projections onto such spaces, is also symmetry under on-site $u(g)$. Moreover, the Hamiltonian remains gapped and TI. In this way, we have found a smooth path of symmetric, in particular TI, Hamiltonian whose adiabatic evolution connects any symmetric state labeled by $\alpha(g)$ and $\om$ to the corresponding fixed point state(up to a local change of basis) and hence establishing the symmetric TI LU equivalence between them.

As we show in appendix \ref{equiv} that fixed point states
with the same $\alpha(g)$ and $\om$ can be related by
symmetric local unitary transformations to each other, we
now complete the proof that \emph{for 1D spin systems with only translation and
an on-site linear symmetry $G$, all gapped
phases that do not break the two symmetries are classified
by a pair $(\om,\al)$ where $\om\in H^2(G,\C)$ label
different types of projective representations of $G$ and
$\al$ label different 1D representations of $G$.}

Similarly, if the system has translation and parity
symmetry, we can establish the equivalence between states
labeled by the same $\alpha(P)$ and $\beta(P)$ in a
translational invariant way (see the discussion below \eq{PA_A}).
The procedure is totally analogous to the on-site symmetry
case, with the only difference that the symmetry condition
for the matrices and double tensors become
\begin{equation*}
\begin{array}{l}
\sum_j u_{ij}A^T_j = \pm M^{-1}A_iM \\ \nonumber
\mathbb{E}_{\beta\chi,\alpha\gamma}=\sum_{\alpha'\beta'\gamma'\chi'} M^{-1}_{\alpha\alpha'} M_{\beta\beta'} (M^*)^{-1}_{\gamma\gamma'} M^*_{\chi\chi'}\mathbb{E}_{\alpha'\gamma',\beta'\chi'}
\end{array}
\end{equation*}
We find that \emph{
for 1D spin systems with only translation and
parity symmetry, there are four gapped
phases that do not break the two symmetries.
}

\end{document}